# Network Resources for Astronomers

Heinz Andernach
Observatoire de Lyon, 9 avenue Charles André, F-69561 Saint-Genis-Laval Cedex, France
heinz@adel.univ-lyon1.fr

Robert J. Hanisch
Space Telescope Science Institute, 3700 San Martin Drive, Baltimore, MD 21218
hanisch@stsci.edu

Fionn Murtagh[1]
Space Telescope – European Coordinating Facility, European Southern Observatory,
Karl-Schwarzschild-Str. 2, D-85748 Garching, Germany
fmurtagh@eso.org

**Running Head:** Network Resources.

**Send proofs to:** Heinz Andernach.

[1] Affiliated to Astrophysics Division, Space Science Department, European Space Agency

## Contents

## ABSTRACT

The amount of data produced by large observational facilities and space missions has led to the archiving and on-line accessibility of much of this data, available to the entire astronomical community. This allows a much wider multi-frequency approach to astronomical research than previously possible. Here we provide an overview of these services, and give a basic description of their contents and possibilities for accessing them. Apart from services providing observational data, many of those providing general information, e.g. on addresses, bibliographies, software etc. are also described. The field is rapidly growing with improved network technology, and our attempt to keep the report as complete and up-to-date as possible will inevitably be outdated shortly. We will endeavor to maintain an updated version of this document on-line.

## 1. Introduction

With the growth of wide-area networking services such as electronic mail, file transfers, information services (bibliographic information, software libraries, etc.), as well as on-line catalogs and databases, astronomers have access to a wealth of information via the terminal or workstation on their desks. This report provides an overview of many of these services, most of them available for free via network connection.

The present report arose from documentation written separately by Hanisch (1992), Andernach (1993) and Feigelson and Murtagh (1992). Some general references on the subject include Jaschek and Heintz (1981); Jaschek (1989); Murtagh and Heck (1988); Albrecht and Egret (1991); Heck and Murtagh (1992); Davenhall (1993b, 1993c); Heck and Murtagh (1993); and Crabtree, Hanisch and Barnes (1994a). While further information on many of the services presented here may be found in these texts, we have made extensive use of information presented in "gray" literature like newsletters, observatory reports, and electronic circulars. An excellent overview of network resources and services, not limited to astronomy, is the book by Krol (1992).

Our report is certainly not complete, and likely biased somewhat by our own interests. Moreover, the information presented in this report is subject to rapid change. For this reason, an on-line version will incorporate changes and corrections. (It will be available by anonymous FTP and other ways at sites `stsci.edu` and `ecf.hq.eso.org`.)

## 2. Network Services

### 2.1. Electronic Mail

Electronic mail (e-mail) is a widely-used service. One needs to know an exact address for one's correspondent. Different forms of address may be feasible due to aliases for machine names which are supported at the correspondent's machine. However an unsupported form of address, or a mistyping of the address, will in most cases lead to a returned ("bounced") message, and in some cases the failure to arrive at the intended destination will not be indicated to the sender. A number of suggestions are given in §7. if you wish to check up on someone's e-mail address. Use of any one of a range of mail programs (e.g. `mail`, `xmh`) is possible, which (still) most often support the transfer of plain text files only. To transfer binary data by e-mail, one should first use some appropriate ASCII coding of the binary data; examples include `uuencode` and `uudecode`.

In this article we employ both Internet and NSI-DECNet (formerly SPAN) address conventions. The DECNet form, `NODE::NAME`, can often be addressed using an Internet routing of the form `name@node.dnet.nasa.gov`. SPAN was closely related to VAX VMS systems, and is being overtaken by universal acceptance of the Internet. Note that the existence of a DECNet address does not necessarily imply that a corresponding Internet address also exists, although it is usually possible to route e-mail via gateway and forwarding systems. The NASA Science Internet Network Applications Information Center (NSI NAIC) provides a very useful guide for e-mail users (contact `naic@nasa.gov`).

#### *2.1.1. Listservers*

"Listservers" are software packages for the support of e-mail distribution lists. They also allow for storage, and automatic access, of files. Although originally developed for use on IBM's Bitnet, listserver software is now available for many systems. There are tens of thousands of listserver distribution lists. Generally messages will be automatically sent to all subscribers if addressed to a list called `xxx-L` at address `xxx-L@node-name.domain`. Subscription and unsubscription requests are dealt with automatically by addressing `listserv@node-name.domain` or, in other implementations, `xxx-request@node-name.domain`. For example, a one-line e-mail message consisting of the text subscribe xxx-L <your_name> results in your e-mail address being added to the distribution list automatically. To check on files stored at the listserver, the one-line message `help` or `index` sent to `listserv@node-name.bitnet` will get you started.

## 2.2. Usenet

Usenet is a peer-to-peer network which came into being in late 1979. For installations which do not rely on telephone line links between machines, it has been superseded by the Internet. By 1981 Usenet was supporting a large volume of newsgroups, and this functionality has been carried over to the Internet. As of August 1994, an estimated 190,000 host machines have access to over 3100 newsgroups and about 7.1 million users at these sites are newsreaders. Not all sites receive the Usenet news, as this requires a site to find another site or organization that will agree to forward news (i.e., act as a "feed"). The access software available to you on your system to access the newsgroup may be a command such as `vnews, rn, trn, xrn, nn, GNUS, Gnews, notes, NNTP`, etc.

A number of newsgroups are devoted to astronomical discussions, and many more are devoted to subjects in physics, chemistry, and computer science that might also be of interest to astronomers. Usenet sites include many of the universities, government research institutions, and observatories where most astronomers work. If you are at one of these sites, Usenet news is probably available to you. If not, you might want to look for someone who would allow you to access news on their system, or you might want to consider becoming a Usenet site yourself. For information about the latter, send e-mail to `mail-server@pit-manager.mit.edu` containing the message

```
send usenet/news.announce.newusers/How_to_become_a_USENET_site
```

Alternatively, this is available by anonymous FTP (see below in this section for further details on anonymous FTP) at the node `pit-manager.mit.edu` in the file

```
pub/usenet/news.announce.newusers/How_to_become_a_USENET_site
```

One key to effective use of Usenet is finding the right newsgroups. Newsgroups are arranged hierarchically by subject of interest. The top level "big seven" hierarchies are `comp`, `sci`, `news` (maintenance issues), `misc`, `rec` (recreational topics), `soc` (social), and `talk` (debate-oriented). Your system may not have all groups, since it is up to individual sites to decide which groups to carry. You may also find additional hierarchies for local or regional groups as well as hierarchies that are not really subject to Usenet "rules", including `alt` (eclectic), `bionet`, `biz`, `gnu` (Free Software Foundation GNU project), and `vmsnet` (VAX/VMS topics). There are more than 8000 newsgroups in the hierarchies mentioned here. For the big seven hierarchies, there is a formal mechanism for creating new newsgroups, and new groups appear regularly.

Of obvious interest to astronomers are the following newsgroups:

| | |
|---|---|
| `sci.astro` | general astronomy discussions and information |
| `sci.astro.fits` | issues related to the FITS image storage standard |
| `sci.astro.hubble` | issues related to the Hubble Space Telescope |
| `sci.astro.research` | discussions of current professional research topics |
| `alt.sci.astro.aips` | discussions on the AIPS image processing system |
| `alt.sci.astro.figaro` | discussions on FIGARO software |

`sci.astro` provides considerable discussion of amateur and popular astronomy questions, but it is read by many professional astronomers and is one possible forum for technical questions.

Other potentially useful newsgroups in the "sci" hierarchy include `sci.chem`, `sci.image.processing`, `sci.optics`, `sci.physics`, `sci.space`, and `sci.space.news`. There are many groups in the "comp" hierarchy that may be of interest, such as the `comp.sys` groups for various vendors of computer hardware, the `comp.os` group for your operating system, the `comp.windows` group for your windowing system (X11, Open Look, Motif, etc.), and if you are a programmer, the `comp.lang` group for the language of your choice. There are newsgroups for software that you might use in your research or for writing papers, such as: `comp.graphics`, `comp.graphics.visualization`, `comp.emacs`, `comp.text.frame`, `comp.text.tex,` and `sci.math.symbolic`.

A nice spin-off of Usenet newsgroups is that many individuals have taken it upon themselves to maintain FAQ (Frequently Asked Questions) files. Since these are often posted to the newsgroup approximately monthly, a good way to see if a FAQ exists is to subscribe to the newsgroup and "listen in." As a condensation of the collective wisdom of the newsgroup (and admittedly signal-to-noise often leaves much to be desired) FAQs provide excellent all-round introductions to a topic, and comprehensive background information.

The "adass" news hierarchy (for Astronomical Data Analysis Software and Systems) has been established in parallel to the Usenet news. This set of news groups is intended as a forum for discussion of astronomical data analysis software. These newsgroups were established by the IRAF Group at NOAO. Information about how to subscribe to this hierarchy, or how to set up a feed, can be obtained by sending e-mail to the Usenet news administrator at `news@tucana.tuc.noao.edu`. The hierarchy currently covers the ADASS Conferences and discussions about IRAF-related software, but other groups are encouraged to establish additional sub-groups.

## 2.3. FTP and Anonymous FTP

FTP ("file transfer protocol") allows copying of files (both ASCII and binary) from a remote machine. A very common practice is to place files of general interest in a restricted-access account which allows "anonymous" FTP access. The details of using anonymous FTP depend a little on the type of system used, but it consists usually of a command such as `ftp remote_system_node_name` followed by a login as user `anonymous` (or as `ftp`). The password is arbitrary, but network etiquette is to provide the user's e-mail address as the password (and some anonymous FTP installations require this!). This allows the provider of the FTP archive to track usage more easily and contact users should that be necessary.

Anonymous FTP allows restricted access to the contents of certain directories. Often files which are intended for general access are available in subdirectories under a `pub` directory. For transfer of files, the commands `get` and `put` are used (transfer of binary files requires setting FTP into binary mode using the command `bin` *prior to initiating the transfer*). In this text we shall refer to relevant data files available via anonymous FTP with nomenclature which combines node name and directory address: `node:dir/file`. By this we mean that one first has to connect to `node` by `ftp node`, then change to directory `dir` (`cd dir`), and finally issue the command `get file` to receive a copy of the file.

## 2.4. Archie

Automatic polling of registered anonymous FTP sites is carried out by a software system called "Archie". Searching through the directories so produced can be a valuable first step, once one knows a program name, in finding a site from which to download the program or file. If Archie has been set up on your system, then a command such as `xarchie` will provide a window interface to some site which supports Archie. To check up on what Archie offers, you may (i) check out some site supporting it, if your site does not; (ii) refer to Emtage (1993); or (iii) send e-mail to `info@bunyip.com`. Registering an anonymous FTP address for search by Archie is done by e-mailing the registration request to `archie-admin@bunyip.com`. NCSA provides a WWW server for access to Archie: use the URL `http://hoohoo.ncsa.uiuc.edu/archie.html` or `http://hoohoo.ncsa.uiuc.edu/cgi-bin/AA`. (See §2.7. for more information about the World-Wide Web and the Mosaic user interface.)

## 2.5. Internet and DECNet Node Names and Addresses

The Internet is actually a collection of networks that are all connected together. Systems on the Internet use a communications protocol known as TCP/IP (Transmission Control Protocol/Internet Protocol). The network resources described in this article generally require access to the Internet, which allows for remote logins and file transfers. Many astronomers are connected via the NASA Science Internet (NSI), or other organizations which provide support for both TCP/IP and DECNet communications. TCP/IP numeric addresses consist of four numbers separated by decimal points (e.g., 130.167.1.2). DECNet numeric addresses are sometimes given as two numbers separated by a decimal point, or as a single number (the convention used here). A DECNet address such as 6.405 is converted to a single number by multiplying the first number by 1024 and adding the second number ($6 \times 1024 + 405 = 6549$).

## 2.6. Remote Logins

The primary mechanisms for establishing an interactive session on a remote computer are `telnet` on systems running TCP/IP and `SET HOST` on systems running DECNet. In either case, in order to use the resources that require an interactive login one needs to know a username on the remote system and, in some cases, a password.

## 2.7. Information Discovery Tools: WAIS, Gopher, and the World-Wide Web

A number of books describing the Internet and the many tools available for locating and retrieving information have recently been published. A survey of these, including intended readership, price, etc. is available in a document by J. Quarterman. This survey is available as RFC 1432. RFC, in the tradition of technical, network-related documents, stands for "Request for Comments". One location among many where this file can be found is `nisc.jvnc.net:pub/RFC/rfc1432.txt`.

In the last two years or so, a number of so-called resource discovery tools have become widely used. Examples are WAIS (Wide Area Information Servers), Gopher, and WWW (World Wide Web). Some of these offer a command-line access mechanism on a machine to which you can login with telnet. However it is much more efficient to use the windowing potentials of such software. As examples, Lynx is a line-mode browser for WWW, and Mosaic is a widely-available window-based browser. Generally, the client-server paradigm is utilized: software sits on your machine, which handles your mouse- or key-controlled requests in your local windowing environment. This *client* communicates over the network with the remote *server*, as the need arises. There is compatibility between these tools, and often Mosaic (for example) is used to access remote sites using Gopher and to carry out free-text searches using WAIS. We will briefly describe these different tools.

Gopher offers a convenient way of accessing anonymous FTP accounts, connecting to remote sites, or receiving files. To use it, and assuming that the client software has been installed at your site, give the command `xgopher`. (This command may be different at your site.) Without a network address following this command, by default you will access a University of Illinois Gopher service. Navigation to various other sites is accomplished by clicking on the appropriate line in what you are shown. Further information on Gopher, including where to obtain client and server source code, is available in Anklesaria and McCahill (1993). A search tool which can be used in conjunction with Gopher is called Veronica.

A WAIS server offers an indexed set of files, so that full- and free-text retrievals may be carried out. A large number of source files (with extension `.src`) which specify the network access addresses of servers are in a Directory of Servers, available via anonymous FTP from `think.com:wais/wais-sources.tar.Z`. You or your system manager should retrieve all relevant source files. Further information on WAIS, including where to find source code for server and client ends of this system, is available in Fullton (1993, 1994). WAIS permits you to index, and thereafter search using free text, your own local text files or mail files. In fact, a wide range of file types can be used as input to WAIS. To display files which are not plain text files (Postscript files, for example) WAIS allows the specification of a filter program which permits display. Thus, if WAIS is informed at the indexing stage that the input files being indexed are of type `PS`, the information may be used subsequently by client software to direct retrieved files through the Ghostview X-terminal screen previewer. To register a WAIS-indexed collection of documents with the WAIS Directory of Servers, one uses the "register" parameter when indexing (which sends the source file to the appropriate registration addresses: `wais-directory-of-servers@cnidr.org` and `wais-directory-of-servers@quake.think.com`).

The WWW (World-Wide Web) is a cross-network hypertext system. It is often used via the user interface Mosaic of NCSA (National Center for Supercomputing Applications). The command `mosaic` (or `Mosaic` or `xmosaic`) may be the one available on your system for access to "the web." A standard form of address and directory information, used by Mosaic, is referred to as a URL (Universal Resource Locator). There are now over 1000 URLs of potential interest to astronomers. Mosaic and the WWW are extremely useful tools for navigating the network, especially because they provide access to other network facilities (Gopher, Archie, FTP, telnet, etc.).

Using a URL via an appropriate browser such as Mosaic gives access to text and graphics, which often encompasses hypertext. This means that some text or graphical icons are "clickable," and gives direct access to a new window of textual or image information. Hypertext is supported by a language (in appearance, not unlike a SGML or Standard Generalized Markup Language, or even LaTeX) called HTML or Hypertext Markup Language. Further details of Mosaic, including source codes for various workstation, PC, and Macintosh platforms, is available at NCSA: `ftp.ncsa.uiuc.edu` in directory `/Web/xmosaic`. A primer on HTML can be accessed at URL `http://www.ncsa.uiuc.edu/General/Internet/WWW/HTMLPrimer.html`.

An informal group of WWW enthusiasts, the AstroWeb Consortium, has been established to maintain a list of URLs of astronomical importance. Members include R. Jackson (CSC/ST ScI), D. Wells (NRAO), A. Koekemoer (Mt. Stromlo), D. Egret and A. Heck (Strasbourg), and H.-M. Adorf and F. Murtagh (ST–ECF). At the URL `http://fits.cv.nrao.edu/www/astronomy.html` these astronomical resources are grouped into the following categories: Observing Resources, Data Resources, Organizations, Software Resources, Publication-Related Resources, People-Related Resources, Various Lists of Astronomy Resources, Astronomical Imagery, and Miscellaneous Resources. A partial list of astronomical URLs is given in Appendix A.

## 3. Astronomical Catalogs and On-Line Databases

A good initial reference for on-line services is Albrecht and Egret (1991) and a new version of this book is currently nearing completion.

A distinction is made in this paper between

- Astronomical catalogs, i.e., static, final compilations of data for a given set of cosmic objects. They can be further subclassified into observational catalogs, compilation catalogs, critical compilation catalogs and bibliographic compilation catalogs (see Jaschek 1989).
- On-line search facilities for published data which allow searches of astronomical objects by name or position on the sky. Examples are SIMBAD, NED and LEDA.
- Observation logs and/or archives of raw or calibrated data of astronomical observatories, like HST, La Palma, WSRT, VLA etc. which can be checked on-line if certain objects have been observed already and if the data are public or not.

The different on-line services described below offer one or more of these items and they are listed under what they predominantly support. The institutions providing these services are often referred to as "data centers". One may broadly speak of large, general purpose data centers (like NASA-NSSDC, of which the "Astronomical Data Center" (ADC) is just a part, and the CDS Strasbourg), medium-sized regional data centers (e.g. in Moscow, Tokyo, La Plata, Beijing, Pune, etc.) and mission-oriented data centers like IPAC, ESRIN, ESTEC, ST ScI, and many others. The latter have mainly been set up in recent times due to the enormous amount of data produced by space missions.

### 3.1. Astronomical Catalogs

NASA and CDS Strasbourg were the first institutions to systematically collect machine-readable versions of astronomical catalogs. These two centers are very comprehensive for ground-based astronomical catalogs, and they have a common agreement to mutually exchange their newly acquired data sets, including a common 4-digit numbering system. However, many other interesting catalogs are not available from these two centers [see e.g. Crézé (1992) or Andernach (1994b)]. Some of them can be found within various other databases (e.g., DIRA2, EINLINE, EXOSAT, HEASARC, STARCAT, STARLINK). Bulk tabular data published in ApJ, ApJS, AJ and PASP are issued on CD-ROM (see Abt 1993), while those published in A&A or A&AS are stored by CDS (see A&A 280, 1 or A&AS 103, 1).

### *3.1.1. NASA Space Science Data Center*

The NASA Space Science Data Center (NSSDC) supports several on-line indices and database systems, including requests for data from the IUE archive and from the Astronomical Data Center (ADC), through its *On-Line Data and Information Service* NODIS. Access is via telnet to `nssdca.gsfc.nasa.gov` (128.183.36.23) or NSI/DECnet to node `NSSDCA` (15548), login as `nodis`. The user will be asked to register and then be presented with a self-explanatory menu which e.g. allows to browse the list of available astronomical catalogs.

The NSSDC has developed an automated data retrieval request service utilizing the "NSSDC Data Archive and Distribution Service" (NDADS). The data held by NSSDC have been written to optical disks which are jukebox-accessed. The NDADS "Automated Retrieval Mail System" (ARMS) permits researchers to rapidly retrieve selections from the current NDADS holdings. Requests are submitted via electronic mail, and the data may be retrieved via anonymous FTP or NSI-DECNet copy. It is also possible to have the data sent directly to the requester's computer. For more information on ARMS, send an empty e-mail message to `archives@ndadsa.gsfc.nasa.gov` or `NDADSA::ARCHIVES` with the words `SEND INFORMATION` in the subject line.

NODIS also provides access to NASA's "Master Catalog" (NMC) which contains information about spacecraft, their instruments, and the data sets therefrom. By summer 1993 data on 4600 orbiting spacecraft and some 60 prelaunch spacecraft were available.

The NSSDC issues a newsletter (on paper) 3-4 times per year which informs about new data sets available from there (including CD-ROMs), mainly from space missions, and including planetary or geophysical data. To subscribe to "NSSDC News", send e-mail to `NCF::REQUEST` and say so. An electronic newsletter of the NSSDC-ADC is also issued and can be subscribed to by sending e-mail to `listserv@hypatia.gsfc.nasa.gov` with one line in the body of the message: SUBSCRIBE ADCNEWS <your_name>.

### *3.1.2. Centre de Données Astronomiques de Strasbourg*

CDS (Centre de Données Astronomiques de Strasbourg) is the oldest European astronomical data center (founded 1972). Like NASA-ADC, the CDS archives some 900 astronomical catalogs (comprising a total of ∼3 Gbyte of data, not including very large data sets such as the IRAS sky maps or the Green Bank radio survey images). Access to most catalogs is offered via anonymous FTP to `cdsarc.u-strasbg.fr` (130.79.128.5) in the subdirectory `/pub/cats`. CDS has recently started to archive files in subdirectories directly named after their published location, e.g., `/pub/cats/J/A+AS/90/327` has tabular data published in A&AS Vol. 90, p. 327. As of June 1994 there are 250 such "J-files". As all other catalogs, these are listed in the file cats.lis. Get the file ReadMe for further information. For example, to retrieve catalog VII/140 (which is the same as A7140 at NASA-ADC), type cd pub/cats/VII/140 and then mget * to copy all files related to that catalog to your computer. Note that some files in the CDS archive (those ending in ".Z") are compressed and that such files should be retrieved with FTP's binary mode. Use the UNIX command uncompress <file_name> to convert them to plain ASCII. For questions or requests for catalogs listed in cats.lis, but not yet available in the FTP archive contact SIMBAD::QUESTION or question@simbad.u-strasbg.fr. CDS produces a regular journal, the Bulletin d'Information du CDS (BICDS), which informs about new catalog acquisitions and related items. A regular e-mail circular with a listing of new catalog acquisitions can be ordered from cats@simbad.u-strasbg.fr or copied directly from file cdsarc.u-strasbg.fr:pub/cats/cats.new. CDS has recently defined documentation standards for catalogs (Ochsenbein 1994) which can be found in cdsarc.u-strasbg.fr:pub/cats/doc.tex. [References: Ochsenbein (1993a,b); Egret and Ochsenbein (1994).]

### *3.1.3. Canadian Astronomy Data Center*

The Canadian Astronomy Data Center (CADC), located at Dominion Astrophysical Observatory (DAO, Victoria) was formed in the mid-eighties to serve as the Canadian center for distribution of data from the Hubble Space Telescope. CADC receives a copy of the HST data in the same way as ST-ECF does (see §3.3.7.). CADC also archives data from the Canada-France-Hawaii Telescope (CFHT, see §3.3.1.). Access to both of these archives is available through STARCAT (see §3.3.7.) and CADC provides its own collection of astronomical catalogs which can be accessed through STARCAT. [Reference: Crabtree et al. (1994b).]

### *3.1.4. Pune, India*

The services of a recently created astronomical data center at Pune (India) can be accessed by telnet to node 144.16.31.6; login as user data. The center currently provides astronomical catalogs which it has obtained by exchange from NASA or CDS, as well as connections to foreign databases for users in India who have no telnet facility. The center plans to archive image data in the future. The catalogs are mostly available on disk, and can be accessed using local software. Send any comments or questions to Ms. Geeta at adc@iucaa.ernet.in or to Ajit Kembhavi (akk@iucaa.ernet.in).

### *3.1.5. CD-ROM Catalogs and Databases*

In 1991 NASA released a CD-ROM with the 114 most requested astronomical catalogs. A list of errata is available via anonymous FTP from hypatia.gsfc.nasa.gov:pub/cdrom/adc_cdrom.errata, or by sending e-mail to listserv@hypatia.gsfc.nasa.gov, with the one line in the body of the message: get pub adc_cdrom.errata.

Numerous other CD-ROMs have been issued, e.g., the Guide Star Catalog (2 CDs), and a CD with sample scans from the ST ScI plate scans of the Palomar Sky Survey. At the time of writing this paper the first set of a total of 101 CD-ROMs with digitized optical surveys of the full sky is being distributed by ASP. There are 31 different CD-ROMs with X-ray data from the EINSTEIN satellite, and four CD-ROMs containing the IRAS high ecliptic latitude Sky Survey Images. Two CD-ROMs with radio images and source catalogs have been issued by NRAO. The "PGC-ROM" with the Principal Galaxy Catalogue (Paturel, Bottinelli and Gouguenheim 1993) includes stand-alone search software. There is one CD-ROM each on "ROSAT, The Images", and on "The Extreme Ultraviolet Explorer Science Archive". A 24-volume set of CDs was issued by NASA-GSFC in September 1992 with data from the Comet Halley Archive. It has already been mentioned that the bulk of the tabular data published in the main North American journals is issued on CD-ROM regularly. The NSSDC On-Line Service (see §3.1.1.) maintains a list of CD-ROMs of astronomical interest, and other compilations have been given by Davenhall (1993b) and within the MediaTheque project of CDS (Heck 1994a).

## 3.2. On-Line Databases

### *3.2.1. Atomic Databases*

There are extensive spectroscopic and other atomic or molecular data in computer-readable files in various laboratories, but the proportion of these that are network-accessible is still small. A bibliography of Atomic Data through to June 1992 was given by Butler (1993), and reviews of the "gray" literature on atomic data have recently been given by Martin (1992a,1992b) and Smith (1993). These papers contain many useful addresses and references.

The Atomic Energy Levels Data Center and the Data Center on Atomic Transition Probabilities at the National Institute of Standards and Technology (NIST) are building an atomic spectroscopic database for astronomy. This database includes energy-level, wavelength, and transition probability data mainly fom the more recent compilations published by the two NIST Centers, as described in sections II and III of Smith and Wiese (1992). These are planned to be made available through the NASA Astrophysics Data System (§3.4.1.).

Two collections of atomic data are available at the Queen's University, Belfast, one on electron impact excitation of atoms and ions, the other on photo-absorption data for atoms and ions, the so-called "opacity project" (OP). The latter collection is more relevant to astronomy, containing, for example, oscillator strengths. The OP is a collaborative project involving institutions in the USA, France, Venezuela and Germany as well as the UK. It comprises approximately 1 GB of data held as ASCII text files. The resulting so-called TOPBASE database is experimentally available from 1993 at CDS Strasbourg. Look in the directory `cdsarc.u-strasbg.fr:pub/topbase` (130.79.128.5). A `Read.Me` file and a user's manual should be consulted before actually using the database through telnet to `cdsarc.u-strasbg.fr`, username `topbase`, password `Seaton+`. For further information contact the CDS staff at `question@simbad.u-strasbg.fr` or Keith Berrington, Department of Applied Mathematics, Queen's University of Belfast, Belfast, BT7 1NN (`amg0016@uk.ac.queens-belfast.app-maths.vax1`). [References: Cunto et al. (1993a,b).]

### *3.2.2. DIRA2*

DIRA2 (Distributed Information Retrieval from Astronomical files) is an ongoing project to manage data from astronomical catalogs being carried out by the ASTRONET Database Working Group in Bologna, Italy. The DIRA2 database contains about 150 original catalogs (∼1 Gb of data) of Galactic and extragalactic data written in a standardized DIRA-specific ASCII format. Several of these catalogs are not available from the NASA-ADC or CDS data centers. Access to DIRA2 is via telnet to `bodira.bo.cnr.it` (137.204.51.8) or via DECNet by `SET HOST BODIRA` (37927). Login as user `dira2` with password `dira2`. Then type `dira2` again to invoke the software. Choose `DB_INFO` and then type `LIST` to see which catalogs are available. The command `WHERE` indicates in which directory they are stored, in case a direct DECNet or FTP copy is preferred (which must be executed from outside the DIRA environment). The output of the searches are ASCII files that can be used in other application programs. Catalogs can also be read or written in the FITS table format. A VT100 terminal emulator is needed for standard searches, and a graphics terminal emulator (Tek4010, VT125) for the graphics tasks. The latter allows one to plot objects in an area of sky taken from various catalogs onto the screen with various symbols of the user's choice. Sorting, as well as selecting and cross-identification of objects from different catalogs is possible. Unix versions for Dec-Ultrix and Alpha OSF/1 are now available and SUN-OS versions are in progress. A reduced version for personal computers ("PC-DIRA") is also available, but lacks many of the graphical routines. During 1993 DIRA2 supported about 1500 remote logins, 40% of these from outside Italy. For the DIRA2 manual as well as for more information, contact M. Nanni (`nanni@astbo1.bo.cnr.it`, `ASTBO1::NANNI`) at the Istituto di Radioastronomia in Bologna, Italy. [References: Benacchio (1991), Benacchio and Nanni (1992), Nanni (1992), Benacchio (1994), Nanni and Tinarelli (1993).]

### *3.2.3. EINLINE*

The Einstein On-line Service (EOLS, or EINLINE) at the Harvard-Smithsonian Center for Astrophysics (CfA) provides access to over 100 data sets (catalogs and mission logs). Many are related to the Einstein X-ray satellite archive, but catalogs of stars, quasars, radio sources (see below), supernova remnants, as well as ROSAT observations can be found. Documentation and ASCII files generated by user queries can be automatically e-mailed to users, and binary data files (such as FITS images from the Einstein CD-ROM collection) can be retrieved using anonymous FTP. There are 31 CD-ROMs available for downloading images and/or photon (list) arrays. EINLINE allows positional searches through groups of similar data sets (all X-ray, all radio, or even ALL catalogs) in one run. This is the MQQ (multiple quick query) option. As a rule, EOLS catalogs are also available through NASA's Astrophysics Data System (ADS, see §3.4.1.).

A special feature of EOLS is the large number of searchable radio-source catalogs collected and provided by the first of the present authors in the course of his activities for the IAU Working Group on Radioastronomical Databases (Comm. 40). As of 1994 August, some 61 source catalogs with ~520,000 entries are searchable, either individually or in affinity groups. This exceeds by several times the amount of radio-source data available from any other on-line service, and most of the tables are not (yet) archived at the established data centers. Some 10,000 lines of documentation have been prepared by the EOLS team, but further submissions of data must be accompanied by appropriate documentation provided by the authors to guarantee their incorporation into EOLS.

Log on to EOLS via telnet to `einline.harvard.edu` (128.103.40.204), via `SET HOST EINLINE` (6714) or else via the WWW under URL `http://hea-www.harvard.edu/einline/einline.html`. Login as user `einline`; no password is required. Documentation for the catalogs can be e-mailed directly from within the EOLS or downloaded with FTP from the directories `einline.harvard.edu:Doc/...`. Documents for any of the catalogs can serve as templates for new contributions. For questions, contact D. Harris (`harris@cfa.harvard.edu`) or C. Stern Grant (`stern@cfa.harvard.edu`). [References: HEAO Newsletters (subscribe to `edpo@head-cfa.harvard.edu`), Andernach (1992), Harris et al. (1992), Andernach (1994a),

Andernach et al. 1994.]

### *3.2.4. EXOSAT*

The EXOSAT database and archive provides on-line access to the results and data products (spectra, images and light curves) from the EXOSAT mission as well as access to data and logs from other missions (e.g., EINSTEIN, COS-B, ROSAT and IRAS). The system is mission-independent and includes timing, image processing and spectral analysis packages as well as software to allow transfer of analysis results and products to the user's home institute. In addition, some familiar optical, IR, and X-ray catalogs, including the HST Guide Star Catalog, are available. EXOSAT allows statistical studies to be performed on large samples of astronomical objects and to retrieve scientific and bibliographic information on single sources. The complete database is located at the EXOSAT observatory at ESTEC, and is accessible via `SET HOST EXOSA0` (29343), `EXOSAT` (28703), or `ESISX0` (29304), or via telnet to `exosat.estec.esa.nl`. Preregistration is necessary: username and address are requested during the first login. Send inquiries to `request@exosat.estec.esa.nl`. [References: White and Giommi (1991), Osborne (1992), Reynolds and Parmar (1993).]

### *3.2.5. HEADS-Brera On-Line Service*

The HEADS-Brera On-Line Service (HEADS = High Energy Astrophysics Database Service) is maintained at the Astronomical Observatory of Brera, Milano, Italy. It is based on the EXOSAT Database System currently developed mainly by NASA's HEASARC (§3.2.6.). HEADS currently provides access to data from a number of past observatories (e.g., Einstein and EXOSAT) and will gradually be extended to include the public data archives of current and upcoming missions. The service also provides access to catalogs, results and data products (images, spectra, light curves) and dedicated analysis packages. There is considerable overlap in the data available through HEASARC, HEADS, EXOSAT, and Leicester (UK), but HEADS supports the complete set of data products (with the exclusion of the ROSAT data, that, in any case, can be obtained from HEADS through the HEASARC database). Login is either via `SET HOST ASTMIB` (32469) or telnet to `astmib.mi.astro.it` (192.167.37.2) and login as user `xray`; no password is required. For a quick reference guide or further inquiries contact either L. Stella (`stella@astmim.mi.astro.it`) or G. Tagliaferri (`tagliaferri@astmim.mi.astro.it`). [Reference: Tagliaferri and Stella (1993).]

### *3.2.6. HEASARC*

The High Energy Astrophysics Science Archive Research Center, HEASARC, is located at the NASA Goddard Space Flight Center in Greenbelt, MD (USA) and its activity is a joint effort between the Laboratory for High Energy Astrophysics (LHEA) and the National Space Science Data Center (NSSDC). HEASARC was created in 1990 to archive and provide access to data from high energy astrophysics missions. As of June 1994 some 125 databases were available. Most of the databases contain data from high-energy astrophysics missions such as EXOSAT, Einstein, Ginga, GRO (Compton), HEAO-1, ROSAT, ASCA (formerly ASTRO-D), BBXRT and XTE, while others contain well-known ground-based catalogs (see Tyler (1994) for a list). Users can sort, search, plot database entries, and cross-identify between these data sets. Telnet to `legacy.gsfc.nasa.gov` (128.183.8.233) and login as user `xray` (no password).

For help on batch-processing of large lists of coordinates, names, and other time consuming queries, send an empty e-mail message to `hdbreq@legacy.gsfc.nasa.gov`. First-time users will be asked to give a HEASARC username, which will be used to allocate a user directory where files can be kept for use in subsequent sessions. Similar services are offered at three other sites in Europe: HEADS-Brera (see above), ESTEC in The Netherlands (see §3.2.4.), and Leicester (for UK users only). The HEASARC also publishes a newsletter called "Legacy"; contact K. Smale at `ksmale@lheavx.gsfc.nasa.gov` to be added to the mailing list. [References: Tyler (1994).]

### *3.2.7. LEDA*

The Lyon–Meudon Extragalactic Database (LEDA) was created in 1983 at Lyon Observatory and is the oldest extragalactic database (Paturel et al. 1988; Paturel, Gouguenheim, and Bottinelli 1992). It gives access to many parameters (up to 66 per object) of astrophysical interest for about 97,000 galaxies, as of June 1994. Finder charts with SAO stars can be created and images taken from the POSS are available for ∼58,000 galaxies in Postscript format. An X-Windows interface allows users to display these on their screen. The main idea is to collect raw measurements from literature and to calculate mean homogenized data in the same spirit of de Vaucouleurs et al. when preparing the series of Reference Catalogs of Galaxies (RC1, RC2, RC3). In fact, the RC3 (de Vaucouleurs et al. 1991) and the Principle Galaxy Catalog (PGC, Paturel et al. 1989; Paturel, Bottinelli, and Gouguenheim 1993) were created using LEDA.

An SQL-like query language allows the user to define and extract galaxy samples by complex criteria. Finder charts in Postscript format can be created with LEDA galaxies and SAO stars of any user-specified position at almost any scale (including that of the POSS and ESO/SERC sky surveys). A batch mode allows queries via e-mail by sending a list of names or coordinates to `ledamail@lmc.univ-lyon1.fr`. Depending on whether the subject specified is `LIST` or `LISTALL`, the system will return (via e-mail) either the basic or all stored astrophysical parameters of the retrieved objects. For subjects `FLAMEQ`, `FLAMGA`, and `FLAMSG` the system will return Flamsteed all-sky projections in Postscript format for any set of input objects (or positions in RA and DEC of B1950).

LEDA is accessible via telnet to `lmc.univ-lyon1.fr` (134.214.4.7), login as `leda`. No password is required. The novice user may choose the menu option `Instructions for Use`. New facilities are frequently added to the system. Suggestions and questions should be sent to G. Paturel (`patu@adel.univ-lyon1.fr`) or M.-C. Marthinet (`mc@lmc.univ-lyon1.fr`).

### *3.2.8. NED*

The NASA/IPAC Extragalactic Database, NED, supported by the Infrared Processing and Analysis Center (IPAC), is a computer-based central archive intended to accumulate a broad range of published extragalactic data (including radio, optical, UV, IR and X-ray sources), and organize them for fast and flexible retrieval via electronic networks. It is currently the most used of such retrieval services. Batch searches for source lists with up to 3000 objects per job are supported, and the results can be retrieved via anonymous FTP (normally within a few hours, after the user is notified via e-mail).

As of 1994 August, NED provides positions, names, and basic data for ∼320,000 extragalactic objects as well as 500,000 bibliographic references to 25,000 published papers and 650,000 photometric data points from catalogs and papers. Also available on-line are abstracts of 9,000 recent articles from several journals (A&A, AJ, ApJ, Ast. Reports, IAU Circ., MNRAS, PAS Japan, PASP) and of 1,000 doctoral dissertations (since 1983) of extragalactic interest. NED encourages submission of English abstracts of any Ph. D. thesis on extragalactic objects. NED has an X Window Graphical User Interface (GUI) supporting mouse-driven menus, and image-display capabilities. Style menus for VT100 terminals continue to be supported. They provide a faster (though more limited) character-based interface. To access NED, telnet to `ned.ipac.caltech.edu` (134.4.10.119) and login as user `ned`. For help or more information, send mail to `ned@ipac.caltech.edu`.

NED is also accessible using Mosaic. If you are familiar with the Web, connect to IPAC using URL `http://www.ipac.caltech.edu/home.html`. [Reference: Helou (1991).]

### *3.2.9. SIMBAD*

SIMBAD (Set of Identifications, Measurements, and Bibliography for Astronomical Data) is produced and maintained by the Centre de Données Astronomiques de Strasbourg (CDS). Access to SIMBAD requires a password, and application may be made by e-mail to `question@simbad.u-strasbg.fr` or `SIMBAD::QUESTION`. SIMBAD may be accessed in the United States via the Smithsonian Astrophysical Observatory (SAO), where a gateway to SIMBAD is maintained under contract with NASA. U.S. astronomers should apply directly to SAO for access, by e-mail to `SIMBAD@cfa.harvard.edu`. SIMBAD charges for its services, but these costs are supported by NASA for astronomers at institutions in the United States. SIMBAD is accessible via telnet to `simbad.u-strasbg.fr` (130.79.128.4) on the Internet or via `SET HOST SIMBAD` (29588). As of March 1994 it contained 1,030,000 objects. For stars (over 665,000 entries) the data include coordinates, spectral type, blue and visual magnitudes, and proper motions. The data for galaxies and quasars (some 80,000 entries) and other non-stellar objects (∼250,000 entries) include coordinates, blue and visual integrated magnitudes, morphological type, size and position angle. In addition there are observational data for some 20 different types of measurements. Over 1,000,000 citations of objects are provided, based on 76,000 bibliographic references complete back to 1950 for stars and to 1983 for non-stellar objects. The total number of identifiers (including those for the PPM star catalog) is 3,300,000. Stars from the Guide Star Catalog can be extracted for any sky region within SIMBAD using the `findgsc` command. As with NED, batch jobs for retrieval of many objects at a time are supported. [References: Egret, Wenger and Dubois (1991), Egret (1992), SIMBAD III (1992).]

### *3.2.10. STARLINK*

STARLINK of Rutherford Appleton Laboratory (UK) offers the Starlink Catalog Access and Reporting System (SCAR), which is a relational database management system for astronomical catalogs. The SCAR package will soon be replaced by CATPAC, described in Starlink User Note SUN 120 by A. Wood (see STARLINK Bull. 10, Oct. 1992). Both packages have the capability to extract data from the requested catalog using input criteria, to manipulate it using various statistical and plotting routines, to output data from the catalog, to assimilate new catalogs and to cross-identify between catalogs based on user-defined criteria (see Starlink User Notes SUN 106 and SUN/70.11). To work on these catalogs one would need to copy the (binary) data file plus a description file within the ADAM environment. However, not all astronomical catalogs exist in the required SCAR-compatible format. The most recent list of available catalogs, as well as information on how to obtain access to a STARLINK account, is available from A. Wood (arw@star.rl.ac.uk). [References: Davenhall (1991), Starlink Bulletins (subscription requests to Mike Lawden at mdl@star.rl.ac.uk), Davenhall (1993b).]

## 3.3. Observatory Archives of Raw or Calibrated Data

There are several observatories which maintain archives of their observational data. Most of these have to be consulted "manually", i.e., by personal request to some archive manager, although many institutions are making their archival data sets available on-line. Here we mention only observational facilities whose data (or at least observation logs) can be accessed or browsed remotely.

### *3.3.1. Canada–France–Hawaii Telescope*

From September 1992 data from several instruments at the Canada–France–Hawaii Telescope (CFHT) have been archived to large optical disks. The CADC is bringing older CFHT data into the archive and by September 1994 the archive will contain data from February 1989 onward. FITS headers of all data frames are sent electronically to the Canadian Astronomical Data Center (CADC) in Victoria daily where an observation catalog is maintained. The catalog is browsable within the STARCAT system and on-line images of public CFHT data can be displayed via the "Preview" system developed by the CADC within Starcat (see §3.3.7.). The proprietary period for CFHT data is two years. [References: Durand et al. (1994), Crabtree et al. (1994b).]

### *3.3.2. Cosmic Background Explorer*

The Cosmic Background Explorer (COBE) was launched in November 1989 and has led to the discovery of fluctuations in the microwave background, as announced in April 1992. The COBE Proposer Information Package gives details of the data sets that will be available. Briefly, a limited set of data (Galactic plane data for FIRAS and DIRBE and the first year's full sky mapping for the DMR) were made available to outside investigators starting in June 1993. The complete data set will be made available in June 1994 in a variety of forms:

- FITS binary tables, available at the COBE data center and through the NSSDC
- FITS image extensions for certain data sets, available at the data center
- COBE-format files at the data center

A Proposer Information Package is available for those who plan to analyze COBE data in response to an Astrophysics Data Program NASA Research Announcement. In addition to general advice for prospective COBE Guest Investigators it contains descriptions of each of the COBE Project Data Sets to be released in June 1994. The package is available as the (Postscript) files nssdca.gsfc.nasa.gov:cobe/project_data_sets/proposer_info*.

There are several non-COBE data sets also kept on line at the COBE analysis center (e.g., IRAS data), some of which have been reprojected and reformatted so that they "look like" COBE data as regards sky projection and resolution. Some of the COBE data have become available through NASA's Space Sciences Data Center (NSSDC). See the file `nssdca.gsfc.nasa.gov:cobe/aareadme.doc` for instructions on how to access such data via anonymous FTP. For request of data on tape contact `request@nssdca.gsfc.nasa.gov`, and for further assistance e-mail to `leisawitz@stars.gsfc.nasa.gov`. [References: White and Mather (1991), Isaacman (1992).]

### *3.3.3. European Southern Observatory*

The archive of ESO observations currently (December 1993) contains around 35,000 images. All observations performed with the ESO NTT (New Technology Telescope, including EMMI and SUSI) since 1 April 1991 are archived. It is accessible through STARCAT (§3.3.7.). [References: Albrecht and Benvenuti (1994), Péron, Albrecht, and Grosbøl (1994).]

### *3.3.4. Extreme Ultraviolet Explorer*

The Extreme Ultraviolet Explorer (EUVE), launched in June 1992, includes a spectrometer and three scanning survey telescopes. The spectrometer samples the range from 7 to 76 nm in three channels (short, medium, and long) with an average spectral resolution of 260. An additional "deep survey channel" co-aligned with the spectrometer provides positional information on observed targets. Two CD-ROMs have been released with EUVE data (S. Bowyer and R. F. Malina, ed.). For more information contact send e-mail to `archive@cea.berkeley.edu` with the word `help` in the body of the message, or browse the anonymous FTP directory `cea-ftp.cea.berkeley.edu:pub/archive`. [References: Christian, Dobson, and Malina (1992); Stroozas et al. (1994).]

### *3.3.5. Compton Gamma Ray Observatory (CGRO)*

Data and information from NASA's *Compton Gamma Ray Observatory (CGRO)* are available from the GRO Science Support Center by telnet to `antwrp.gsfc.nasa.gov`, login as user `gof`. Among the information offered there is e.g. the 2B catalog of 585 gamma ray bursters from the *Burst and Transient Source Experiment* (BATSE). Time profiles for all events are available via WWW (`http://enemy.gsfc.nasa.gov/cossc/cossc.html`). [Reference: Hanlon 1994.]

### *3.3.6. HIPPARCOS*

HIPPARCOS (High-Precision Parallax Collecting Satellite) is an ESA satellite which was launched in August 1989 and is dedicated to the measurement (at the milliarcsecond level) of positions, trigonometric parallaxes, and proper motions for about 120,000 stars. The target list (the HIPPARCOS Input Catalog) is available from CDS or NASA–ADC as catalog I/191 or A1191, respectively (see §3.1.1. and §3.1.2.). A series of papers on the satellite performance and first results appeared in A&A 258 (1992). For further information contact C. Turon (turon@mesioa.obspm.fr) or D. Morin (morin@mehipa.obspm.fr). [References: Turon et al. (1991, 1992a, 1992b).]

### *3.3.7. Hubble Space Telescope*

Both the ST ScI and the ST-ECF, as well as the CADC (see §3.1.3.) provide on-line information services and access to the complete HST Archive. The HST data archive is described in Pasian et al. (1993) and Long et al. (1994). As of mid-1994 the archive is comprised of some 2,000,000 files in 400,000 data sets and occupies 1.1 TB of storage space on optical disks, and new data continues to be archived at an average rate of 1.5 GB/day. Subject to proprietary period restrictions (typically one year) preview images of HST data are accessible through both the StarView user interface at ST ScI and through the STARCAT user interface at the ST–ECF and the CADC. The "preview data" are generated at the CADC and then sent to both the ST-ECF and ST ScI.

ST ScI runs a system called Space Telescope Electronic Information Service (STEIS), which is an anonymous FTP archive (also accessible via gopher and WWW) containing information on HST proposals, instruments, observation logs, and software. Of particular interest is the Archive Exposure Catalog (AEC), a list of all completed observations. Copies of the AEC reside in directory `hst-archive` in the files `README.AEC`, `AEC.CATALOG`, and `AEC_SS.CATALOG`, the latter containing solar system observations. A listserver (`listserv@stsci.edu`) supports many types of current awareness services, such as instrument news, weekly and daily summaries of performance, planned observing time-lines, and lists of completed observations. ST ScI's WWW home page is located at `http://stsci.edu/top.html`.

ST ScI has two host computers providing external access to the HST archive: `stdata.stsci.edu` (130.167.1.135, VMS operating system) and `stdatu.stsci.edu` (130.167.1.148, Unix operating system). To get started with the HST archives, telnet to one of these machines and login as user `guest` with password `archive`. Run the Starview user interface by entering the command `starview` (for the simple ASCII terminal version) or `xstarview` for the X Windows version. Note that you can also access the HST archive via ST ScI's WWW home page.

In order to actually retrieve data, the user must register (type `register` when logged in on `stdata` or `stdatu`, or register via the WWW archive page). Each user is given an individual account on either stdata or stdatu (user's choice). Data retrieved from the archive will appear in the `data` directory of `stdata` or `stdatu`. Users are sent e-mail when all requested data has been staged to these areas, and then the data can be transferred to your home machine using FTP. Use the command `readnews` to be kept up to date on the status of the archive and changes to the retrieval software.

Similar services are provided by the ST–ECF for European users of HST. The bulletin board service STINFO is accessed by telnet to `stinfo.hq.eso.org` with login username `stinfo` (no password). Questions about HST can be sent to an e-mail hotline `stdesk@eso.org` or `ESO::STDESK`. An anonymous FTP account, which includes a software library, various other software items, documentation, test images, and so on, is at address `ecf.hq.eso.org`. If you have WWW access, then the quickest on-line source of information regarding the the archive for HST and ESO data is available using the URL `http://arch-http.hq.eso.org/ESO-ECF-Archive.html`.

The ST–ECF also distributes HST data to European archival researchers. Data requests are prepared from within STARCAT. While catalog browsing is anonymous, a registration is necessary for retrieving data.

The ST–ECF copy of the HST archive is available via telnet to `stesis.hq.eso.org`; login as user `starcat`, or contact `catalog@eso.org` for help. STARCAT also provides an interface to a number of astronomical catalogs (listed in the "Documentation for on-line catalogs" item), including the capability of sorting, selecting, and cross-correlating catalogs. On-line versions of the "Star*s Family" of directories, dictionaries, and data sets (see §7.1. and §7.2.) are available from within STARCAT. For further information and user's guides, contact B. Pirenne at the ST–ECF (`bpirenne@eso.org`, `ESO::BPIRENNE`). For documentation on STARCAT send a request to `shill@eso.org`.

Canadian users should request HST archive data through the CADC; contact `cadc@dao.nrc.ca` for more information.

[References: HST Catalog User Guide (1992); Schreier, Benvenuti, and Pasian (1991); Ochsenbein (1991); Pirenne et al. (1992); Adorf, Jackson, and Murtagh (1993); ST–ECF Newsletter (subscription requests to `rfosbury@eso.org`); ST ScI Newsletter (subscription requests to `elliott@stsci.edu`).]

### *3.3.8. Infrared Processing and Analysis Center*

IPAC (Infrared Processing and Analysis Center, Pasadena, CA, USA) offers on-line access to its datasets and services. IPAC is a primary data node for the ADS (§3.4.1.), and provides several object catalogs which can be accessed independently (telnet to `xcatscan.ipac.caltech.edu` and login as user `xcatscan`). This service requires the user to run an X Windows server. Contact Rick Ebert `rick@ipac.caltech.edu` for further details. IPAC issues a Newsletter on paper; subscription requests, as well as general questions about network services at IPAC, should be sent to `info@ipac.caltech.edu`. IPAC supports the NASA/IPAC Extragalactic Database (see §3.2.8.) as well as the *ISSA Postage Stamp Service* (see §3.3.9.). [Reference: Ebert (1994).]

### *3.3.9. IRAS*

A full set of uncalibrated IRAS (InfraRed Astronomical Satellite) data is stored on an optical disk jukebox at the Space Research Center in Groningen, accessible semi-online through a mail server. The server provides data extraction, calibration, and imaging for survey and Low Resolution Spectral Data. Analysis software (GEISHA) and display software (GIPSY) are used. More interactive access to the database is planned for the future. The system is transportable to other sites. The manual for remote access to the IRAS server can be obtained via e-mail to `irasman@sron.rug.nl`. An FTP connection allows retrieval of data. Data can be requested by sending e-mail to `iras_server@sron.rug.nl`. The request can select the data for the area of interest, recalibrate fluxes and position of the infrared measurements, and recombine them into an image. Several options can be set for this processing to obtain the best possible reconstruction of the IR sky for the user's research. The user will be notified when the request is processed, and about how to retrieve the data by FTP. Note that before using the mail server, the user has to pre-register by sending e-mail to `irasman@sron.rug.nl`. For further questions contact D. Kester, IRAS manager (`do@sron.rug.nl` or `irasman@sron.rug.nl`).

The experimental *ISSA Postage Stamp Service* is available via WWW (`http://brando.ipac.caltech.edu:8888/ISSA-PS`) and allows to retrieve maps of size 2×2 degrees of a user-specified area in all four IRAS bands (12, 25, 60 and 125 microns) with a pixel size of 1.5 arcmin.

The maps will be displayed on the screen and can also be retrieved in FITS format for further display and processing.

The IRAS Low Resolution Spectra are available via telnet to `hyades.colorado.edu`, login as user `lrsuser`. A more complete set of spectra is available from the University of Calgary. Users should send e-mail to `kwok@iras.ucalgary.ca` for more information. The 11,000 spectra from the University of Calgary are also available from the CADC via STARCAT (§3.1.3.), including on-line views of the spectra.

Four CD-ROMs with the IRAS Sky Survey Atlas Images were released by IPAC in mid-1992, one each for the three sky coverages HCON-1, HCON-2 and HCON-3, and another one for the average of them. For updated overviews of IRAS data products in general, see a recent issue of the IPAC Newsletter (§3.3.8.). [References: Walker (1991); Roelfsema, Kester, and Wesselius (1993); Wesselius et al. (1992).]

### *3.3.10. Infrared Space Observatory*

ISO (Infrared Space Observatory) is an ESA space observatory to be launched in September 1995. Up-to-date information on the mission can be obtained via anonymous FTP from `ftp.estec.esa.nl:pub/iso` or `ftp.ipac.caltech.edu:pub/iso`. Information on ISO on the WWW is being set up under URLs `http://isosa2.estec.esa.nl:8223/` and `http://www.ipac.caltech.edu`. Data centers are being established at Heidelberg (Germany) and Rutherford Appleton Labs (RAL, UK); see, e.g., Gabriel et al. (1992). The *ISO/IRAS Newsletter* is edited by H. Walker (`hjw@star.rl.ac.uk`), and *ISO Info* is a newsletter issued by ESA (contact K. Leech at `kleech@isosa6.estec.esa.nl`).

### *3.3.11. International Ultraviolet Explorer*

IUE (International Ultraviolet Explorer), a satellite launched in January 1978 and still producing data, is equipped with a 45-cm telescope and two spectrographs for the range 1150 to 3200 Å, with resolutions of either 300 or 20,000. IUE led to the first modern astronomical data archive, and is also the first heavily used archive. Currently, all data requested from the archive corresponds to over 50 years of observing time, i.e. almost four times the actual mission duration. Three centers maintain a complete up-to-date copy of the IUE Archive: the NASA IUE archive at NSSDC (GSFC, USA), the ESA IUE archive at VILSPA (Spain), and the SERC IUE archive at RAL (UK). The plan is to have all IUE data remotely accessible in a final archive by 1994 (De La Peña et al. 1994). IUE Newsletter 41 (Oct. 1992) presents a merged observation log from 1978–1991 on microfiche. The observing log can be queried on-line: `SET HOST VILSPA` (28843), login as user `VILSPA` with password DB. The user is provided with a menu. Detailed usage of the system is described in the ESA IUE Newsletter 37 and an example of the scientific potential of the archive — when used systematically — is given by laDous (1994). The IUE Uniform Low Dispersion Archive (ULDA) is a uniform subset of the IUE archive, and is distributed by ESA/Vilspa. ULDA is available through national hosts, and further information may be obtained from A. Talavera (`VILSPA::AT` or `iuehot@esoc.bitnet`). The ESA IUE Newsletter is published quarterly by ESA and NASA (contact `iueobs@v3300.vilspa.esa.es` or `VILSPA::IUEOBS`). [References: Wamsteker (1991), Ponz et al. (1992), laDous (1993).]

### *3.3.12. James Clerk Maxwell Telescope Archive*

The Royal Observatory Edinburgh (ROE) maintains a data archive of observations made with the millimeter-wave James Clerk Maxwell Telescope (JCMT) in Hawaii. The archive contains data obtained since January 1992. Data become public after a proprietary period of one year. The observation log is maintained on-line and can be interrogated remotely using the ARCQUERY software (see text on La Palma (§3.3.13.) and Westerbork (§3.3.15.) below). Access is via `SET HOST RLESIS` (19527); login as `ESIS`, select menu option `1`, answer `REVAD`, and login as `ARCQUERY`. Alternatively, one can telnet directly to host `star.roe.ac.uk` and login as `arcquery`. On-line documentation is available via telnet to `star.roe.ac.uk`, login as `jcmtinform`. Copies of a User's Guide and Quick Reference Sheet can be obtained from A. McLachlan at ROE (`aml@star.roe.ac.uk`). The JCMT Newsletter is edited by G. Watt at ROE (`gdw@star.roe.ac.uk`). [References: Davenhall (1993a), Hummel and Davenhall (1993).]

### *3.3.13. La Palma Archive*

This archive contains most of the raw telescope data taken with the Isaac Newton Group (ING) of optical telescopes on La Palma (the 1-m Kapteyn (JKT), 2.5-m Newton (INT), and 4.2-m Herschel (WHT) telescopes). The data are physically stored on tapes at RGO (Royal Greenwich Observatory, Cambridge, UK). The observation log has 260,000 entries as of April 1994 and is searchable on-line through the ARCQUERY software at Cambridge (UK): `SET HOST RLESIS` (19527), login as `ESIS`, select menu option `1`, answer `CAVAD`, and login as `ARCQUERY`. Alternatively, telnet to `gxvg.ast.cam.ac.uk` (131.111.69.20) and login as `arcquery`. For further information contact E. Zuiderwijk at `ejz@mail.ast.cam.ac.uk`. [References: Raimond and Van Diepen (1989), Raimond (1991, 1994), Zuiderwijk et al. (1994).]

### *3.3.14. ROSAT*

ROSAT (Röntgensatellit), lauched in mid-1990, is a collaborative effort of Germany, the UK, and the US, and provides imaging, spectral, and timing information on sources in the extreme UV and soft X-ray wavebands. ROSAT has performed an all-sky X-ray survey with an estimated number of 60,000 detected sources. Pointed observations are ongoing, from which information also on ~20,000 serendipitous sources per year is expected. Three centers (NASA–HEASARC, see §3.2.6., LDS Leicester, and MPE Garching) provide access to observation logs and to the public data (after a proprietary period of one year).

Access to ROSAT data and other ROSAT-related information is available by anonymous FTP to `rosat_svc.mpe-garching.mpg.de`. The directory `archive/data` contains observations ordered by identifier.

Access to the ROSAT archive for Starlink users is also available from the Leicester Database System (LDS) via DECNet: `SET HOST RLESIS` (19527), login as `ESIS`, select option `1`, answer `LTXDB`, and login as `XRAY`. A data request form is available from this node via DECNet as the file `LTVAD` (19838)`::DISK$ROSAT:[ROLOC.PUBLIC.ARCHIVE]REQUEST.FORM`. A "User Guide for the UK ROSAT Data Archive" is available via anonymous FTP from `darc.star.le.ac.uk:rosat/user_guide.tex`. Contact S. Sembay (`sse@star.le.ac.uk`) or M. Watson (`mgw@star.le.ac.uk`).

Data from ROSAT's Wide Field Camera (WFC) are available from the Space Data Center at RAL (UK); see D. Giaretta and E. Dunford (1991). [References: Zimmermann and Harris (1991), Zimmermann (1992), Paul (1992); Voges (1992), Zimmermann et al. (1992), Sembay and Watson (1992), UK ROSAT Electronic Newsletter (subscription requests to `julo@star.le.ac.uk`).]

### *3.3.15. Westerbork Synthesis Radio Telescope Archive*

This archive contains all of the raw data ever taken with the Westerbork Synthesis Radio Telescope (WSRT), an aperture synthesis telescope of 14 antennas of 25-m diameter in The Netherlands. The data are physically stored on tapes at the Netherlands Foundation for Radio Astrononomy (NFRA), Dwingeloo, The Netherlands. The observation log contains 78,000 entries as of April 1993 and is searchable on-line with the same software, ARCQUERY, as for the La Palma archive (see §3.3.13.). Access is via telnet to `rzmvx1.nfra.nl` (192.87.1.100), login as `arcquery`. For further information contact E. Raimond at (`exr@nfra.nl`). [References: Raimond (1991, 1994), Zuiderwijk et al. (1994).]

### *3.3.16. Lunar and Planetary Institute*

The databases of the Lunar and Planetary Institute (LPI) are available by telnet to `cass.jsc.nasa.gov` (192.101.147.17), login as `cass` with password `online`. The LPI Center for Information and Research Services provides resources on geology, geophysics, astronomy, and astrophysics. The available files include catalogs of journal holdings, books, and maps, a bibliography of the lunar and planetary literature from 1980 on, and an index to the Benchmarks in Geology Series. The system is menu driven. For comments and questions contact D. Bigwood (`bigwood@lpi.jsc.nasa.gov`). Also note that ESA issues the "Solar System News," edited by the Planetary and Space Science Division (PSS). Contact K. Wenzel at `kwenzel@estec.bitnet`.

### *3.3.17. Solar Data*

The National Geophysical Data Center (NGDC) offers numerous data sets on diskette or CD-ROM (contact `info@mail.ngdc.noaa.gov`). The two most extensive on-line databases of solar data are the Space Environment Laboratory (SEL) database and the Solar Terrestrial Dispatch (STD) service. These are comprised mostly of optical data, geophysical and solar indices, flare reports, X-ray activity, and so on. The STD also contains at least one radio datum, the 10.7-cm flux value. For further information, contact D. Speich `dspeich@selvax.sel.bldrdoc.gov` (SEL) or C. Oler, `oler@rho.uleth.ca` (STD). Solar Maximum Mission (SMM) data are available from the Space Data Center (SDC) at Rutherford Appleton Laboratoires (RAL, UK), and various NASA solar data sets are available via the NSSDC (§3.1.1.). Data are stored off-line and copied by special request onto disk, to be transferred to a user's home computer. The electronic Newsletter "SolarNews" is published monthly and can be subscribed to by sending e-mail to `SOLAR` (25000) `::EDITOR`. The European Physical Society issues the "European Solar Physics Newsletter" once or twice per year; contact R. J. Rutten in Utrecht at `rutten@ruunsc.fys.ruu.nl` or `SOLAR::RRUTTEN`. ESA issues the "Solar System News," edited partly by the Solar and Heliospheric Division (SHS). Contact A. Pedersen at `apederse@estec.bitnet`. The Japanese Data Center "Solar-Terrestrial Environment Laboratory" (STELAB) issues the "STEP GBRSC NEWS" (send subscription requests to `gbsrc@stelab.nagoya-u.ac.jp`).

## 3.4. Distributed Databases

The fact that each data archive has its own software, user interface and storage format makes the use of many different databases difficult for the standard user. Therefore new interfaces have been established which allow uniform access to very different databases stored even in geographically different places. So far two such systems are available: ADS in the United States and ESIS in Europe.

### *3.4.1. Astrophysics Data System*

ADS (Astrophysics Data System) is NASA's principal on-line research facility, and is a research tool that is intended to make access to data and information convenient and easy for the science user. The ADS uses an interface that allows its users to make queries based on a multitude of astronomical data sets that are physically located in different locations, but available for querying in a uniform way, regardless of how the data sets are organized and independent of where they are actually located. This liberates the users from learning different access methods. The ADS includes tools for data visualization once these data have been retrieved. A set of database management functions is available for manipulating the retrieved tables of objects which can be written to files in several formats, including ASCII and FITS. These formats allow data obtained via ADS to be imported into astronomical data analysis systems such as IRAF, AIPS, or MIDAS. The ADS also includes visualization tools that allow users to make 2-D plots of one table column versus another.

Registration forms to access ADS may be obtained from the ADS user support at `ads@cuads.colorado.edu`. Apart from access to many other services which are described separately in the present paper, the ADS Release v4.0 of January 1994, provides access to over 250 catalogs, including over 50 radio source catalogs and over 161,000 abstracts searchable with complex algorithms (also see §5. for access to abstracts via other services).

Further information about ADS is available via their WWW home page (URL `http://adswww.harvard.edu/adswww/adshomepg.html`). A number of ADS services are available directly through the WWW, and access through this method is typically faster, easier to use, and does not require registration. The majority of the catalogs can be accessed at URL `http://adswww.harvard.edu/catalog_service.html`. The ADS Abstract Service is available via the URL `http://adswww.harvard.edu/abstract_service.html`. Other ADS services available via the WWW are listed at URL `http://adswww.harvard.edu/ads_services.html`. [References: Weiss and Good (1991), Giovane (1992), Murray et al. (1992), Kurtz (1993), Kurtz et al. (1994), Eichhorn (1994).]

### *3.4.2. European Space Information System*

ESIS (European Space Information System) aims at providing the user with a set of tools to access, exchange, and handle information from a great variety of sources including space mission archives, databases of scientific results, bibliographical references, white and yellow page directories (also see §7.2.), etc. The ESIS bibliographic service (ESISBIB) allows abstracts from the NASA RECON collection to be retrieved. Different from the ADS abstract servive (see §3.4.1.), ESISBIB uses abstracts from a wider range of other astronomy-related topics (e.g. geophysics, computer sciences, theoretical mathematics, etc.) with altogether ~500,000 abstracts. ESISBIB makes use of the SIMBAD references and automatically links any SIMBAD reference with its abstract, if available in the system. Access to all information on ESIS is possible through a homogeneous query and data managing-language based on a Graphical User Interface (GUI). Version 2.0 was released in February 1994 and client software is available for VAX/VMS and SUN/UNIX machines and is installed at about 40 sites. Access to ESIS is via telnet to `esis.esrin.esa.it` (192.106.252.127 or 192.106.252.100) or `SET HOST ESIS` (29617). Login as `astronomy` with password `evalme`. The user will be required to register on first login, and identify him/herself in subsequent sessions. ESIS is also available on the WWW via the URL `http://mesis.esrin.esa.it/html/esis.html`. Send queries to `ESIS::ISDHELP` or `isdhelp@mail.esrin.esa.it`.

ESIS NEWS is a newsletter to which one may subscribe (send requests to `ESIS::ISDHELP`). It is available also via anonymous FTP as a color Postscript file (`mesis.esrin.esa.it:pub/esis/newsletter/esisnews.ps`). [References: Albrecht (1991, 1992), Giommi and Ansari (1994a,b).]

## 4. Plate Catalogs, Digital Optical Sky Surveys, and Finding Charts

Many relevant contributions on this matter can be found in *Digitized Optical Sky Surveys* (MacGillivray and Thomson 1992), in the newsletters of the IAU Working Group on Wide-field Imaging (contact the editor, H. T. MacGillivray at `hmg@star.roe.ac.uk` for copies), and in the proceedings of the IAU Symposium 161 (MacGillivray et al. 1994).

### 4.1. Plate Catalogs

There are many observatories keeping archives of their plate materials. Lists of such observatories have recently been compiled (Tsvetkov and Tsvetkova 1993, Dluzhnevskaya 1994) and some are available electronically on request to `tsvetkov@bgearn.bitnet`.

#### *4.1.1. The UK Schmidt, ESO Schmidt, and AAT Plate Catalogs*

An extensive photographic plate library is maintained at the Royal Observatory of Edinburgh (ROE). It contains some 15,000 original plates taken with the UK Schmidt Telescope (UKST) in Australia, some 6,500 original plates taken with older telescopes of ROE, and copies of all major photographic sky surveys. A complete catalog of all plates taken with the UK Schmidt Telescope is accessible via DECNet (`SET HOST REVAD` (19889) or `RLESIS::REVAD`) or through telnet to `star.roe.ac.uk`. Login as user `UKSCAT`; no password is required.

Plate catalogs for the Anglo-Australian Telescope (AAT) and ESO Schmidt Telescope are also maintained. (Note that the AAT catalog does not include digitally recorded observations.) To access either of these catalogs proceed as for the UKST catalog, replacing the login name by either `AATCAT` (for the AAT catalog) or `ESOCAT` (for the ESO Schmidt catalog) and follow the instructions. For further information contact ROE staff (`ukstu@star.roe.ac.uk` on Internet or `REVAD` (19889) `::UKSTU` on DECNet).

### 4.2. Digital Optical Sky Surveys

The digitization of several large-scale optical sky surveys (the Palomar and ESO/SERC atlases) has been completed in recent years using high-performance plate-scanning machines. The raw scan data are so bulky that remote access to this data is still prohibitive. However, the object catalogs extracted from these scans are being made accessible over the network. An overview of the various projects is given by MacGillivray and Thomson (1992).

#### *4.2.1. Automated Plate Measuring Machine*

The *Automated Plate Measuring Machine* (APM) is located at the Institute of Astronomy, Cambridge, UK. It has been used to scan the POSS-I survey plates at galactic latitudes above $|b| = 20°$. Both colors were scanned and objects cross-identified, so that color information is available (so far unique in all-sky surveys) for a matched object catalog of well over 100 million objects down to m=22 in blue (O) and m=20 in red (E). At this time, no copies of the entire catalog are distributed, but potential users will be given access to smaller subsets for their own use. The northern hemisphere catalog above $-3°$declination is available for routine interrogation and the southern hemisphere catalog based on UKST $B_j$ and R plates is about half complete and also available. A captive account for remote catalog interrogation can be accessed via telnet to `131.111.68.56`, login as user `catalogues` and follow the instructions. Contact: M. Irwin (`mike@mail.ast.cam.ac.uk` or R. McMahon (`rgm@mail.ast.cam.ac.uk`). [Reference: McMahon and Irwin (1992); Irwin, Maddox, and McMahon (1994).]

#### *4.2.2. COSMOS*

COSMOS (COordinates, Sizes, Magnitudes, Orientations, and Shapes) is a plate scanning machine at the Royal Observatory Edinburgh, which has been used to scan the whole southern sky (DEC $< +2.5°$) from the IIIa-J and Short Red Surveys, and has led to an object catalog of several hundred million objects. During 1994 public access to the catalog will be provided through the Anglo-Australian Observatory (AAO). Retrieve the file `cosmos_catalogue.lis` from the directory `aaoepp2.aao.gov.au:aao_obs` (Drinkwater 1994) or contact M. Drinkwater (`mjd@aaocbnu1.aao.gov.au`). A new digitization of the southern sky survey using the SuperCOSMOS machine will give an even better quality object database, which is planned to be made available on CD-ROM by 1995. Contact H. T. MacGillivray (`hmg@star.roe.ac.uk`). [References: Yentis et al. (1992), MacGillivray (1992).]

#### *4.2.3. ST ScI Guide Star Plate Scans*

Digitizations of both the Quick-V and the POSS E and O plates and ESO-IIIa-F red plates (limiting magnitude is 20 and 22, respectively) are available in a near on-line manner to local ST ScI users or any visitor to ST ScI. Compressed versions of the southern survey scans are available on a set of 60 CD-ROMs since 1994, while the images from the northern POSS-E (red) plates will be distributed on 40 CD-ROMs in early 1995. In addition to this set of 100 CD-ROMs with a compression factor of 10, another set of 10 CD-ROMs will be produced with a compression factor of 100. This latter one (to be available in 1995) will be useful mainly for educational purposes and for amateurs and will be sold by the ASP. For other optical sky surveys being digitized at ST ScI see the list by McLean (1994). [Reference: Lasker (1992); Postman (1994); editorial note in PASP 106, 108 (1994).]

The scanning of the first plates of the POSS-II survey has begun at ST ScI, while the object catalog is being prepared at Caltech (see Weir and Djorgovski 1992, Weir et al. 1994). The proceedings of the IAU Symposium 161, ("Astronomy from Wide-Field Imaging," H. T. MacGillivray et al., 1994) provides more information.

#### *4.2.4. Minnesota Automated Plate Scanner*

A digitization of the POSS-I E- and O-survey plates was also performed with the Automated Plate Scanner (APS) at the Astronomy Department of University of Minnesota. So far all plates (E and O) have been scanned for $|b| > 20°$. All of the images detected on *both* E and O plate are being entered in a catalog with $\sim 10^9$ stellar objects and a few times $10^6$ galaxies. This catalog is available at URL `http://isis.spa.umn.edu/homepage.aps.html`. Object catalogs for some Palomar plates have become available as ADS catalogs (see 3.4.1.). Contact address: C. Cornuelle at `aps@aps1.spa.umn.edu`. [References: Pennington et al. (1992, 1993), Aldering et al. (1994).]

### 4.3. Finding Charts

Finding charts may be used for locating suitable guide stars for a particular telescope, for transparent overlays to find objects on Schmidt survey plates, and for visualization purposes when cross-correlating catalogs. The largest catalog of stars currently available for this purpose is the ST ScI Guide Star Catalog (GSC), which is available on CD-ROM from the ASP. It contains positions which are precise to between $0\overset{''}{.}4$ and $0\overset{''}{.}7$. Note, however, that this catalog is not magnitude-limited, but that the selection of stars has been carried out so as to provide a homogeneous *density* of guide stars over the sky. The EXOSAT and SIMBAD databases (see §3.2.9.) and others (e.g., the GASP package in IRAF/STSDAS, or STELLA, which is described below) allow guide stars to be selected from user-defined areas of the sky. A "pocket version" (300 MB) of the GSC has been produced by Preite-Martinez and Ochsenbein (1994).

SKYMAP is a computer program which produces maps of arbitrary portions of the sky in a variety of projections and coordinate systems. Over the past 10 years it has been used to produce finding charts for very different purposes. It can display multiple source catalogs, including the HST Guide Star Catalog, as well as solar system objects with astrometric accuracy. SKYMAP can be tuned to a specific task using an ASCII parameter file which controls how information is displayed on any Tektronix-compatible graphics display or hardcopy device. The program contains a variety of interactive graphic and image processing features and has been ported to a variety of computer systems. The program is available by anonymous FTP from `cfa0.harvard.edu` (128.103.40.1) in the `pub/gsc` directory. Contact D. Mink `mink@cfa.harvard.edu` for an illustrated manual of the program. [Reference: Mink (1993).]

Other packages to produce finding charts from the GSC have been developed, e.g., by Megevand (1992), Smirnov and Malkov (1993), and Malkov and Smirnov (1994). Finding charts based on the SAO star catalog can be produced with the `CHART` option within the Starlink software package (Allan 1989). SAO star charts as well as images of the surroundings of 58,000 galaxies (from POSS–I) can also be generated using LEDA (see §3.2.7.). "Preview" images of galaxies from the *Surface Photometry Catalogue of the ESO-Uppsala Galaxies* (Lauberts and Valentijn, 1989), are available through the STARCAT interface (see §3.3.7.).

SkyView is a facility available over the net which allows users to retrieve data from public all-sky surveys conveniently. The user enters the position and size of the region desired and the catalogs desired, and the data is extracted and formatted for the user. SkyView handles coordinate conversions and provides the user with data in the desired coordinate system. A description of SkyView is in the SkyView User's Guide and SkyView Design document, both available via anonymous FTP in `skview.gsfc.nasa.gov:pub/doc`. SkyView also has extensive internal help available. Access to SkyView is via telnet to `legacy.gsfc.nasa.gov`, login as user `xray`. No password is required, but when you first log in you will be asked for information to set up your own user file area. To start SkyView, type `skyview` and you will start up the SkyView GUI user interface. An article on SkyView will be appearing in the next issue of Legacy, the Journal of the High Energy Astrophysics Archive Research Center (HEASARC). A scaled down WWW version is available as the URL `http://skview.gsfc.nasa.gov/skyview.html`. For questions or problems with SkyView contact K. Scollick (`scollick@skview.gsfc.nasa.gov`) or T. McGlynn (`mcglynn@grossc.gsfc.nasa.gov`). [Reference: McGlynn, White, and Scollick (1994).]

STELLA is a chart generator supporting access to the GSC, SAO, PPM, IRAS, and other catalogs. These are accessed through STARCAT (see §3.3.7.). STELLA is available to users at ESO (contact B. Pirenne at `bpirenne@eso.org`).

## 5. Bibliographical Services and Preprints

Until recently, bibliographic services were only available on a fee-for-use basis. Overviews of the available services were previously given by Rey-Watson (1988), Watson (1991), and more recently by Davenhall (1993c,d). The commercial or fee-for-use bibliographic databases available via the network include INSPEC, SPIN, PHYS, STN, and Dialog; ideally your local librarian should be the expert in access to and use of these facilities. A number of bibliographic and other information services are available on the net at no charge, however. Some of these are described below.

### 5.1. Bibliographic Services

#### *5.1.1. Project STELAR*

STELAR (STudy of Electronic Literature for Astronomical Research) is a project that NASA has undertaken to develop on-line access to both abstracts and full text for much of the recent astronomical literature. Abstracts are being provided from NASA's RECON system. It provides abstracts back to the mid-1960's, but even for the period covered systematically (1975 onwards) a complete coverage of any of the sources is not guaranteed. The pages of the ApJ, ApJS, AJ, PASP, and A&A are being scanned beginning with issues from 1987, with the plan to make the pages available as bit-maps. The first service to be offered by STELAR is access to journal abstracts, AAS meeting abstracts, and the AAS Job Register. Access is provided by a WAIS (Wide Area Information Service) server. Details are provided in the October 1992 issue of the AAS Newsletter. The project STELAR WWW homepage is `http://hypatia.gsfc.nasa.gov/STELAR_homepage.html`. For more information, send e-mail to `stelar-info@hypatia.gsfc.nasa.gov`. [References: Van Steenberg (1992), Van Steenberg et al. (1992), and Warnock et al. (1992, 1993).]

#### *5.1.2. Astrophysics Data System Abstract Service*

The ADS (see §3.4.1.) abstract database contained 161,557 abstracts as of October 1993. Five astronomically-related categories of the NASA STI database are used to select relevant abstracts. The data include abstracts from all major journals, many minor journals, NASA reports, and many PhD theses. Users can query by author, object name, keywords, words in the title, and words in the abstract text. Logic within and between these fields can be changed, and the importance of a given word is weighted based on the frequency of that word in the database. The resulting list is ranked by how closely the paper matches the query. From this list, one can obtain all of the information contained in the ADS abstract database (including the bibliographic code, publication date, category, title, authors, author affiliations, keywords, and abstract text). Query feedback capability is provided which uses the results of one query to form another.

Collaboration with SIMBAD (see §3.2.9.) has enabled ADS to provide the capability of searching by object name, and references from SIMBAD are correlated with those in the ADS abstract database. Access to the ADS abstract service is available on the WWW at URL `http://adswww.harvard.edu/abstract_service.html`.

#### *5.1.3. Lunar and Planetary Institute*

The Lunar and Planetary Institute offers a bibliographic service which can be accessed by telnet to `cass.jsc.nasa.gov` (192.101.147.17). Login as user `cass` with password `online`.

#### *5.1.4. UnCover*

UnCover is a database of tables of contents of over 14,000 journals, including ApJ, A&A, AJ, ApJS, MNRAS, Nature, PASP, and Science. In late 1993 it included more than 4,000,000 articles and over 1,000,000 articles are added annually. It can be accessed by telnet to `pac.carl.org` or `database.carl.org` and offers keyword searches for article titles or authors. Data ingest started in September 1988 and the first year of complete coverage is 1989. The contents pages of an individual journal issue can be viewed by volume and number, and further details can be obtained on individual articles. Copies of all retrieved articles can be ordered (by FAX only) for a charge indicated by the database. A telnet connection to `pac.carl.org` also allows one to access other library databases and browse the library catalogs of several North American libraries. Send inquiries to `help@carl.org` or `database@carl.org`.

#### *5.1.5. Abstracts and Indices of Astronomical Journals*

The Center for Astrophysics provides on-line access to the contents pages of the ApJ, ApJS, AJ, and PASP, as well as to the abstracts of articles that have been submitted to the ApJ Letters. Both services are available via telnet to `cfaN.harvard.edu` (where N = 3, 4, 5, 7, or 8) or SET HOST to DECNet node `CFAn` (n as above). To review titles of papers login as user `apjaj`. To see the ApJ Letters abstracts, login as user `aplett`. Both logins provide simple instructions on how to use the facilities. PASP abstracts and ToCs are available via anonymous FTP to `stsci.edu:pasp`, via Gopher to `stsci.edu` and on the WWW at `http://cdsweb.u-strasbg.fr/CDS.html`.

Following an agreement with the Editors of *Astronomy & Astrophysics*, the CDS provides access to the abstracts from this journal, starting in January 1994. These abstracts are made available four weeks before the publication of the corresponding issue. The abstracts can be viewed and retrieved on the WWW at URL `http://cdsweb.u-strasbg.fr/Abstract.html`. They are also available through anonymous FTP from the directory `cdsarc.u-strasbg.fr:pub/abstract`. A WAIS server is in preparation. Send remarks and comments to `question@simbad.u-strasbg.fr`.

The NED database (see §3.2.8.) is systematically entering the abstracts of papers of extragalactic interest from five major journals (A&A, AJ, ApJ, MNRAS and PASP) complete since 1988, the IAU Circulars since 1991, and Soviet Astronomy (now Astronomy Reports) and PASJ complete since 1992.

#### *5.1.6. Acta Astronomica Archive*

The journal *Acta Astronomica* (Poland) keeps an electronic archive of its recently published papers. They can be retrieved via anonymous FTP to node `sirius.astrouw.edu.pl` (148.81.8.1). Papers are stored in subdirectories named `acta/year_of_volume/author_page/`, e.g., `acta/1992/kal_29/` where "author" is an abbreviation of the first author's last name, appended with the starting page of the paper (see also Acta Astr. 4, issue 4, Editorial).

### 5.2. Preprint Services

The AstroWeb Consortium (§2.7.) maintains a list of on-line preprint services of interest to astronomers. Some of the major facilities are described in more detail below.

#### *5.2.1. Los Alamos National Laboratory*

A fully automated bulletin board for preprints is available at Los Alamos National Laboratory. To get information on its use, send the word `HELP` in the subject field of a message to `hepth@xxx.lanl.gov`. Information on subscribing, accessing, etc., will be automatically returned. Access by anonymous FTP is also possible. Typically 200–300 papers are available each month.

#### *5.2.2. CERN*

The European Centre for Nuclear Research (Geneva) supports a preprint server. Via WWW use address `http://darssrv1.cern.ch` provides access to hundreds of subdirectories, each associated with one or a few institutes. An X Windows based access mechanism is also supported. Refer to van Herwijnen (1993) for details of this.

### 5.2.3. SISSA

Fully automated bulletin boards for preprints in astronomy and astrophysics are available at SISSA (International School for Advanced Studies), Trieste, Italy. Preprints can be submitted and retrieved. It is also possible to subscribe to a daily distribution list containing authors and abstracts of the papers submitted to the bulletin board. The service is operated by listserver software at the address `astro-ph@babbage.sissa.it` (147.122.1.21). Preprints are mostly in TEX/LATEX format but Postscript format as well as compression and/or encoding is also common. Facilities to use special macros are available, and figures in Postscript format may also be submitted. Anonymous FTP access is also possible. For further details, instructions, disclaimers, etc., send mail with a subject line saying `help` to the address given above. The same host also supports abstract services for other fields of physics.

### 5.2.4. Preprint Lists from NRAO and ST ScI

The libraries of the National Radio Astronomy Observatory (NRAO) and the Space Telescope Science Institute (ST ScI) cooperate in producing lists of preprints received. The lists are distributed electronically every two weeks by the librarians of ST ScI and NRAO (see below). Both preprint lists are also available through WAIS.

STEPsheet ("Space Telescope Exhibited Preprints") is a list of all preprints received during the last two weeks at the ST ScI and is prepared by the ST ScI librarian, Sarah Stevens-Rayburn. It is delivered by electronic mail and subscription requests should be sent to `library@stsci.edu`. Each list presently contains well over 100 titles. Note that the preprints themselves are not distributed by the ST ScI librarian and must be requested from the individual author. The full current ST ScI database contains everything received in the last several years, along with all papers received since 1982 and not yet published. It is searchable on-line by connecting to `stlib.stsci.edu` (130.167.1.175) and logging in as `stlib`. Note that VT100 keypad emulation is required to use the EDT editor for searching. Both the current ST ScI database and the full database of all preprints received since 1982 are searchable via WAIS as `stsci-preprint-db.src` and `stsci-old-preprint-db.src` in the WAIS `directory-of-servers`. These files are also available on WWW from the library's homepage, `http://sesame.stsci.edu/library.html`. For additional information, contact `library@stsci.edu`.

The RAPsheet ("Radio Astronomy Preprints") is a listing of all preprints received in the Charlottesville library of the National Radio Astronomy Observatory in the preceding two weeks. It is meant to serve as an alert service only — the NRAO library does not copy or distribute the preprints listed. Interested persons should request copies of preprints from the authors. The unRAPsheet is a listing of papers which appeared previously on the RAPsheet for which citations have been added in the preceding two weeks. The tables of contents of all incoming journals and meeting proceedings are scanned in order to find citations and update the records. The RAPsheet is posted to the `sci.astro` Usenet group and the full database is available via anonymous FTP in the `pub/rapsheet` directory on `polaris.nrao.edu` and is updated once every two weeks. A database of preprints received, along with their added citations, from 1986 forward, including unpublished ones since 1978, is also searchable using WAIS. The necessary source file is `polaris.nrao.edu:pub/rapsheet/nrao-raps.src`. Please address any questions, comments, or corrections regarding the RAP/unRAPsheets to `library@nrao.edu`.

### 5.3. Astrolib

Astrolib is an e-mail distribution for astronomy librarians whose purpose is to share relevant information widely and rapidly. Astrolib was born out of the desire expressed at IAU Colloquium 110 (Wilkins and Stevens-Rayburn 1988) to continue the valuable exchange of information begun there. Astrolib messages vary widely in content, including information on difficult-to-find and noncommercial items (both print and non-print), information on publication problems or defective journal issues, duplicate items offered to other libraries, reports from conference attendees, reports on ongoing projects (e.g., the IAU Thesaurus, see §7.1.), questions about technical library matters, and information on resource sharing. E. Bouton of the National Radio Astronomy Observatory acts as manager and moderator for the Astrolib distribution. She receives and redistributes all messages, editing them and combining them as necessary or convenient. Information about librarians' e-mail addresses is regularly and actively solicited by the manager and by other librarians. In late 1993 there were 120 people from 23 countries receiving Astrolib messages. For further information contact E. Bouton, `library@nrao.edu`.

### 5.4. On-Line Library Card Catalogs

ST ScI (Space Telescope Science Institute) offers public access to its library catalog. Telnet to `stlib.stsci.edu` and login as user `stlib` (no password) and follow the menu. A user guide is available on request from `library@stsci.edu`.

The contents of the ESO libraries at Garching bei München and La Silla can be browsed remotely. Access is via telnet to `libhost.hq.eso.org`, login as user `library`. Two user guides are available: "The ESO Libraries On-line Catalogue in a Nutshell" and "The ESO User Guide to the On-Line Catalogue." Both are available from the ESO Library in Garching (`esolib@eso.org`). [Reference: ESO Messenger 74, 1993, p. 37.]

The card catalogs of hundreds of libraries (including the Library of Congress, which can be accessed via telnet to `locis.loc.gov`) are available over the net. Look into the directory `ftp.unt.edu:pub/library`. The UnCover system (§5.1.4.) also provides access to several larger US libraries (not specialized in astronomy).

## 6. Electronic Publishing

Introductions to this wide and growing field were recently given by Heck (1992a), Heck (1992b), and in the Newsletter of the American Astronomical Society No. 62 (October 1992), Special Insert. The first and last references discuss the prospects of future directions for electronic publishing, the standards for text editors, page scanning, etc.

Below we list the currently available macro packages for preparation of papers for major astronomical journals.

- *ApJ, AJ, PASP*. These journals encourage the use of the AASTEX package, available via FTP. To obtain details on the retrieval and installation of the current version (3.0), send an empty message to `aastex-instruct@blackhole.aas.org`.

- *A&A*. Send e-mail to `svserv@dhdspri6.bitnet` with either the command `send tex/plain/p-aa.zip` or `send tex/latex/l-aa.zip` in the main body of the message, to receive the TEX or LATEX mark-ups for papers in *Astronomy and Astrophysics*.
- *MNRAS*. Macros for electronic papers to be submitted to *MNRAS* are obtainable via anonymous FTP to `tex.ac.uk`. The TEX files are in `[tex-archive.macros.plain.contrib.mnras]` and the LATEX files are in `[tex-archive.macros.latex.styles.contrib.mnras]`.
- *Ap&SS*. Prospective authors for *Astrophysics and Space Science* should request a TEX macro from the editor, Kluwer Academic Publishers at `editdept@wkap.nl`.

## 7. Dictionaries, Thesauri, Directories, Meetings, and Jobs

### 7.1. Dictionaries and Thesauri

The *Second Dictionary of the Nomenclature of Celestial Objects* has just appeared (Lortet, Borde and Ochsenbein 1994a,b). Authors of survey-type source lists are strongly encouraged to check that designations of their objects do not clash with previous namings and are otherwise commensurate with recommendations on nomenclature. An on-line installation of the *Interactive Dictionary of Acronyms* is provided by telnet to `simbad.u-strasbg.fr` (130.79.128.4); login as `info` (no password). The command `info cati ACR` will then provide information on the over 3000 acronyms `ACR` of which over 700 are also recognized by SIMBAD. The command `info -l cati ACR` provides more details on a given acronym. [Reference: Lortet, Borde, and Ochsenbein (1994a,b).]

The dictionary StarBriefs, maintained by A. Heck (Strasbourg), is available and searchable via WWW under URL `http://cdsweb.u-strasbg.fr/CDS.html`. It currently comprises about 70,000 abbreviations, acronyms, contractions and symbols from astronomy and space sciences, as well as related fields. Astronomers are invited to consult this dictionary to avoid assigning an acronym that has been used previously. [Reference: Heck (1993).]

On behalf of the IAU several librarians of large astronomical institutions have been working since 1986 on the compilation of a thesaurus of astronomical terms. The first version (1.1) was distributed in paper form to various observatories, and is now available electronically in the directory `aaoepp2.aao.gov.au:lib_thesaurus`. There are versions for three different operating systems (MS-DOS, Macintosh and Unix). The thesaurus will be essential in many respects, e.g., to aid authors in better selection of keywords for their papers, and to aid librarians in better classification of publications. It will also be used as a basis for sophisticated bibliographical search algorithms using the full text of papers (Kurtz et al. 1994, also see §3.4.1.).

To allow searching and classification of non-English literature, a French, German, Italian and Spanish version of the primary terms is being worked on. The second draft version of this "Multi-Lingual Supplement to The Astronomy Thesaurus" has been distributed in May 1994. For their own benefit in future literature searches, all astronomers are invited to check the terms in their respective fields of specialization and make suggestions to R. Shobbrook, librarian at AAO (`lib@aaoepp2.aao.gov.au`). [References: Shobbrook and Shobbrook (1992a, 1992b, 1993).]

### 7.2. Directories

Reviews of "Astronomical Directories" were given by Heck (1991) and Heck (1992c).

A few years ago C. Benn and R. Martin undertook the task of collecting e-mail addresses of individual astronomers in a single ASCII table. The table is arranged in five columns (name, institution, network, electronic address, and date code, the latter indicating when update information was last received from a given institution), and updates are released every year. They are available via anonymous FTP in three files located at `ftp.ast.cam.ac.uk:guide/ASTRO*`. The same directory also contains some useful documents on electronic mail and TEX macros to print the RGO directory. This list of individual e-mail addresses is ASTROPERSONS.LIS and the current version of April 1994 contains ∼11,000 entries, of which 92 % are on the Internet. Alternative e-mail addresses for observatories and astronomy departments are given in file ASTROPLACES.LIS. Postal addresses are given in file ASTROPOSTAL.LIS. Requests for copies of these files, corrections, and additional information for inclusion in the next edition can also be sent to `email@mail.ast.cam.ac.uk` or directly to C. Benn `crb@lpve.ing.iac.es`.

The dictionary StarGuides, maintained by A. Heck (Strasbourg), is available and searchable via WWW under URL `http://cdsweb.u-strasbg.fr/CDS.html`. It comprises practical information on associations, societies, scientific committees, agencies, organizations, etc., in astronomy, space sciences, and related fields. Currently about 5000 items from 100 countries are offered, with information on postal and electronic addresses, phone and FAX numbers, staff, activities, geographical coordinates of observing sites, awards or prizes granted, etc. StarGuides supersedes the book *Astronomy, Space Sciences, and Related Organizations of the World (ASpScROW)*. For additional information contact A. Heck (`heck@cdsxb6.u-strasbg.fr`). [References: Heck (1994b).]

The Unix commands `whois` and `netfind` are both valuable tools for tracking down network contacts. Both may be accessed by going to `Phone Books` on most Gopher servers.

### 7.3. Astronomical Meetings

E. Bryson, Librarian at the Canada-France-Hawaii Telescope Corporation in Kamuela, Hawaii has been compiling a list of forthcoming astronomical meetings, to which one can subscribe electronically (by request to `library@cfht.hawaii.edu`). Each mailing has the new meetings marked with an asterisk. Bryson does her best to achieve completeness for the new meetings, but clearly relies on timely information from the meeting organizers. The current listing of astronomy meetings is available via anonymous FTP from the file `ftp.cfht.hawaii.edu:pub/library/meetings.doc` as well as on the WWW at URL `http://cadcwww.dao.nrc.ca/meetings/meetings.html`.

### 7.4. Jobs

The American Astronomical Society offers its monthly list of job openings in astronomy (the "Job Register") by anonymous FTP from the directory `blackhole.aas.org:jobs/jobreg`. Note that usually only the latest month is kept as an on-line file. Contact `ssavoy@blackhole.aas.org` if you want to be on the mailing list.

STARJOBS is an electronic notice board maintained at the Rutherford Appleton Laboratory and sponsored by the European Astronomical Society (EAS). Telnet to `star.rl.ac.uk` or `SET HOST RLSTAR` and login as `starjobs`. You will enter a hierarchical help library at the subtopic called `JOBS`. To read a job notice type the first few characters of the title of the notice and follow with a carriage return. Note that there is no facility for mailing back the notices to you. For local hardcopies it is best to keep a log of your session or cut and paste to extract text. See the EAS Newsletter No. 7 (Jan. 1994) for more details.

Under the URL `http://chronicle.merit.edu/` users can find an electronic listing of academic job openings in all fields, including astronomy. This service is provided by the Chronical of Higher Education and contains primarily but not exclusively openings in the United States.

## 8. Astronomical Software

An article by Feigelson and Murtagh (1992) discusses publicly available astronomical software and numerical libraries and offers additional hints about using the network. An example of one of these services is the node `netlib@research.att.com`, which will send any one of a number of numerical algorithms upon receipt of an e-mail message of the form `send name_of_routine`. Send a message saying `send index` to see what is available. A similar service for statistical software is available from `statlib@temper.stat.cmu.edu` (128.2.241.142).

An Astronomical Software Directory Service has been initiated by the second of the present authors (Hanisch, Payne, and Hayes 1994). The objective of this project is to provide astronomers with the ability to search for certain types of algorithms, applications programs, or utilities, across the many software packages in use in the community. Users will be able to read documentation, learn about software and hardware requirements for the code, and get pointers to anonymous FTP archives where code may be retrieved.

Many of the data analysis packages used in the astronomy community are available in FTP archives. A table of such systems is given below. Note that some groups require that you make previous arrangements with them in order to retrieve their software, although patches are usually freely available.

TABLE 1 GOES HERE

Information about the FITS (Flexible Image Transport System) data format standard is also available on-line. The NSSDC provides the NASA description of the FITS standard on nssdca.gsfc.nasa.gov (128.183.36.23). More FITS-related documents, sample data, test files, and a log of the transactions in the Usenet newsgroup sci.astro.fits are available on fits.cv.nrao.edu (192.33.115.8). Most of the files can also be found on ftphost.hq.eso.org (134.171.8.4). A library of I/O routines for reading and writing FITS files has been developed by Bill Pence at HEASARC; look in the directory pub/fitsio3 on tetra.gsfc.nasa.gov (128.183.8.77), or send e-mail to pence@tetra.gsfc.nasa.gov.

At Penn State University the *Statistical Consulting Center for Astronomy* has recently been created (see the directory ftp.stat.psu.edu:pub/scca or the URL http://www.stat.psu.edu/scca/homepage.html).

Pointers to other software packages of interest to astronomers are listed within the MediaTheque project of CDS; see URL http://cdsweb.u-strasbg.fr/CDS.html (Heck 1994a).

## 9. Other Information Systems and Services

ATNF (Australia Telescope National Facility) maintains the Australia Telescope, a radio interferometer at Narrabri. Observing proposals and documentation on this facility can be obtained via anonymous FTP from files in the directory atnf.csiro.au:pub/atnfdocs. A README file in this directory contains further details.

ESO (European Southern Observatory) has all press releases including accompanying photographs, lists of all preprints, some full preprints, continually updated weather images of South America and Europe, instrument information, and so on available through its WWW URL http://www.hq.eso.org/eso-homepage.html.

IAU Circulars are posted on various electronic bulletin boards. Contact addresses are marsden@cfa.harvard.edu or green@cfa.harvard.edu.

IRAM (Institut de Radioastronomie Millimétrique, Grenoble, France) offers an electronic newsletter containing documentation on the IRAM telescopes on Pico Veleta (Granada, Spain) and the radio interferometer on Plateau de Burre (near Grenoble) and on reduction software, files for proposal preparation, etc. For details use anonymous FTP to retrieve the file dist/newsletter/README on node iraux2.iram.fr (193.48.252.22) or send e-mail to newsserv@iram.fr with only the word HELP in the body of the mail. If you want to be notified when a new newsletter becomes available, send a message to iramusers-request@iram.grenet.fr with the one word SUBSCRIBE in the body of the message. For further questions send e-mail to Robert Lucas at lucas@iram.fr or lucas@iramfr51.bitnet.

The La Palma Observatories (Observatorio del Roque de los Muchachos) distribute an electronic newsletter, the "Isaac Newton Group — La Palma Information Bulletin". Its purpose is to keep the community of users of the optical telescopes on La Palma (Canary Islands, Spain) informed of the actual situation on the mountain top. Distribution is by e-mail only. For subscriptions send your e-mail address to bulletin@ing.iac.es, or, for UK users, to dxc@castro.ast.cam.ac.uk.

NASA Headquarters maintains a WWW server at http://www/mtpe.hq.nasa.gov/HQ_homepage.html. Of particular interest to astronomers are the anonymous FTP directories for Astrophysics, which include the full text of NASA Research Announcements and the newsletters of the Astrophysics Division (the FTP host is ftp.astrophysics.hq.nasa.gov).

NRAO is performing two sky surveys with the VLA. Within the *NRAO VLA Sky Survey* (NVSS) the whole sky north of $-42°$ declination is being mapped at 20 cm. The survey will run from 1993 to 1996 at an angular resolution of $\sim 50''$ and be complete to a flux density limit of 2 mJy. The NVSS is being made as a service to the astronomical community, and the *uv* data and maps are being released via anonymous FTP in the directory `gibbon.cv.nrao.edu:pub/nvss` immediately as they are taken, calibrated, and mapped. Later there will be CD-ROMs and hardcopy listings. An introduction to the NVSS is also available on the WWW at URL address `http://info.aoc.nrao.edu`. Within another survey (FIRST, or *Faint Images of the Radio Sky at Twenty-cm*) the north Galactic cap ($b > +30°$) will be surveyed from 1993 to 2000 at 20 cm at an angular resolution of $\sim 5''$) to a flux density level of 1 mJy. [References: Condon (1994); Condon et al. (1994); Becker, White, and Helfand (1994).]

NOAO (National Optical Astronomy Observatory) maintains several public FTP archives on node `gemini.tuc.noao.edu` (140.252.1.11), including `weather` (weather satellite pictures), `noao` (information about NOAO), and `preprints` (Postscript versions of selected NOAO preprints). Past issues of the NOAO Newsletters are available for searching via WAIS on `pandora.tuc.noao.edu`.

SMTO (Submillimeter Telescope Observatory, Tucson, USA) has commissioned a 10-m telescope for sub-mm observations on Mt. Graham (Arizona) and distributes a newsletter about current progress of the project. Send a message to `mailserv@gaffel.as.arizona.edu` to receive further details on how to obtain these newsletters.

VLAIS (Very Large Array Information System), supported by the National Radio Astronomy Observatory (NRAO, USA), contains information about the VLA radio interferometer in New Mexico and provides access to the observations log of all observations made with the VLA, and also with the VLBA (Very Long Baseline Array), a network of radio antennas distributed over the US. Other general information is available, e.g. a master address list for astronomers, NRAO employees e-mail addresses, etc. To access, telnet to `zia.aoc.nrao.edu` (146.88.1.4) and login as user `vlais`. Instructions are provided on-line. It is possible to e-mail yourself some of the information in the system; follow the instructions under `MAILIT`. Send queries to Carl Bignell (`cbignell@nrao.edu`). [Reference: Bignell (1990).]

Weather information, especially for North America, is provided at many sites. Meteosat and other images of Europe, Australia, North America, etc., are updated regularly during the day. The "Sources of Meteorological Data FAQ" gives information about images and forecasts, and is available from `rtfm.mit.edu:pub/usenet/news.answers/weather-data.Z`. Chilean weather forecasts are available via Gopher at `tortel.dcc.uchile.cl` (146.83.4.40); log in as `gopher`, and then choose successively: "Servicios Miscelaneos", "Pronosticos Meteorologicos", "Informe Diario Direccion Meteorologica de Chile").

E. Bryson, Librarian at the Canada-France-Hawaii Telescope Corporation in Kamuela, Hawaii, is working on a database with a WAIS client that will be an international clearing house of observatory manuals and the address of how to acquire them. Send inquiries to her at `library@cfht.hawaii.edu`.

## 10. Conclusions

This paper may be the last of its kind to be published on paper, as increasing amounts of information are offered via network tools like WAIS and WWW. Relevant information is changing so fast that upon failure to connect using the information given in the present paper, the reader is encouraged to use other pointers from this article to find the desired information.

Many of the descriptions of the various systems were taken from published introductory information provided by the system sponsors. We are grateful to A. C. Davenhall for providing us with his reviews prior to publication and for his careful reading of the manuscript. Many other people have contributed valuable information and suggestions, in alphabetical order: L. Benacchio, P. Boyce, L. Brotzman, E. Bryson, G. Dulk, G. Eichhorn, P. Giommi, D. E. Harris, J. Hayes, A. Heck, G. Helou, F. Hill, K. Long, R. Isaacman, R. Jackson, A. de Jonge, D. Kester, H. T. MacGillivray, M.-C. Marthinet, W. C. Martin, K. Nakajima, M. Nanni, F. Ochsenbein, H. Payne, B. Pirenne, A. Reynolds, M. Schmitz, F. Simien, K. Smale, P. L. Smith, S. Stevens-Rayburn, R. M. Shobbrook, C. Stern Grant, M. Van Steenberg, M. Vazquez, and M. Wenger. We thank the referees, E. D. Feigelson and D. R. Crabtree, as well as the Editor, H. Bond, for pointing out several other useful services to us. H. A. acknowledges a research grant of CNRS and is grateful for the hospitality he received at Instituto de Astrofísica de Canarias and Observatoire de Lyon.

This preprint was prepared with the AAS LaTeX macros v3.0.

## A. URLs of Note

URLs (Uniform Resource Locators) provide the address information needed to make a link with a World-Wide Web site. A popular user interface to the WWW is NCSA Mosaic. A number of lists of astronomical resources have been established, and coordination between those maintaining these lists has been undertaken by a group calling itself the "AstroWeb Consortium". Current members are: Jackson (CSC/STScI), Wells (NRAO), Koekemoer (Mount Stromlo), Egret and Heck (Strasbourg Observatory), and Adorf and Murtagh (ST-ECF). The AstroWeb Consortium has formed a common resource listing, and assembled various tools – in particular, automated submission procedures for new URLs, and automated checking procedures to verify that URLs still respond to accesses as expected. Currently, more than around 1000 astronomically-relevant URLs are available. At URL `http://fits.cv.nrao.edu/www/astronomy.html`, these astronomical resources are categorized under: Observing Resources, Data Resources, Organizations, Software Resources, Publication-Related Resources, People-Related Resources, Various Lists of Astronomy Resources, Astronomical Imagery, and Miscellaneous Resources.

A short list of URLs for sites which support user services — archive research, or observing — is given below. No guarantee can be made that required information will be found at these sites. However, the momentum behind the World-Wide Web is currently very great, and the information available through Mosaic (or other browsers) is growing constantly in quality and quantity. The World-Wide Web is thoroughly compatible with other network-based access mechanisms: WAIS, Gopher, FTP, telnet, etc., and in fact provides a common user interface to all of these.

ADS (NASA Astrophysics Data System)
`http://adswww.harvard.edu/adswww/adshomepg.html`
CADC (Canadian Astronomy Data Center)
`http://cadc.dao.nrc.ca/CADC-homepage.html`
CDS (Strasbourg Data Center)
`http://cdsweb.u-strasbg.fr/CDS.html`
CEA (Center for EUV Astrophysics)
`http://cea-ftp.cea.berkeley.edu/HomePage.html`
CTIO (NOAO Cerro Tololo Interamerican Observatory)
`http://ctios2.ctio.noao.edu/ctio.html`
ESIS (European Space Information System)
`http://mesis.esrin.esa.it/html/esis.html`
ESO (European Southern Observatory)
`http://www.hq.eso.org/eso-homepage.html`
HEASARC (High Energy Astrophysics Science Archive Research Center)
`http://heasarc.gsfc.nasa.gov`
NASA (National Aeronautics and Space Admininstration Information Services)
`http://hypatia.gsfc.nasa.gov/NASA_homepage.html`
Mount Stromlo and Siding Springs
`http://meteor.anu.edu.au/home.html`
NRAO (National Radio Astronomy Observatory)
`http://info.aoc.nrao.edu/`
NOAO (National Optical Astronomy Observatories)
`http://www.noao.edu/noao.html`

NASA (NASA Online Information)
`http://mosaic.larc.nasa.gov/nasaonline/nasaonline.html`
National Solar Observatory
`http://argo.tuc.noao.edu/`
Planetary Data System
`http://starhawk.jpl.nasa.gov/pds_home.html`
ST-ECF (Space Telescope – European Coordinating Facility)
`http://ecf.hq.eso.org/ST-ECF-homepage.html`
ST ScI (Space Telescope Science Institute)
`http://stsci.edu/top.html`

## B. List of Acronyms

| | |
|---|---|
| AAO | Anglo-Australian Observatory |
| AAS | American Astronomical Society |
| ADC | Astronomical Data Center (of NASA's NSSDC) |
| ADS | Astrophysics Data System |
| AEC | Archive Exposure Catalog (of the Hubble Space Telescope) |
| AIPS | Astronomical Image Processing System |
| APM | Automated Plate Measuring Machine (Cambridge, UK) |
| APS | Automated Plate Scanner (Univ. Minnesota, USA) |
| ARMS | Automated Retrieval Mail System (of NASA-ADC) |
| ASCA | Advanced Satellite for Cosmology and Astrophysics (formerly Astro-D) |
| ASCII | American Standard Code for Information Interchange |
| ASP | Astronomical Society of the Pacific |
| ATNF | Australian Telescope National Facility |
| BATSE | Burst and Transient Source Experiment (on board CGRO) |
| BBXRT | Broad Band X-Ray Telescope |
| BICDS | Bulletin d'Information du Centre de Données Astronomiques de Strasbourg |
| Bitnet | Because It's Time network |
| CADC | Canadian Astronomical Data Center (Victoria, BC) |
| CD-ROM | Compact Disk - Read-Only Memory |
| CDS | Centre de Données Astronomiques Strasbourg (Strasbourg Data Center) |
| CERN | European Center for Nuclear Research |
| CfA | Center for Astrophysics |
| CFHT | Canada-France-Hawaii Telescope |
| CGRO | Compton Gamma Ray Observatory |
| COBE | Cosmic Background Explorer |
| COSMOS | Coordinates, Sizes, Magnitudes, Orientations and Shapes (ROE, UK) |
| DEC | Digital Equipment Corporation |
| DIRA2 | Distributed Information Retrieval from Astronomical files |
| DIRBE | Diffuse Infrared Background Experiment (on COBE satellite) |
| DMR | Differential Microwave Radiometer (on COBE satellite) |
| EAS | European Astronomical Society |
| EINLINE | Einstein On-Line Service |
| EMMI | ESO Multimode Instrument |
| EOLS | Einstein On-Line Service |
| ESA | European Space Agency |
| ESIS | European Space Information System |
| ESO | European Southern Observatory |
| ESRIN | European Space Research Institute (ESA) |
| ESTEC | European Space Technolgy and Research Center (ESA) |
| EUVE | Extreme Ultraviolet Explorer |
| EXOSAT | European X-ray Observatory Satellite |
| FAQ | Frequently asked questions |
| FIRAS | Far Infrared Absolute Spectrophotometer (on COBE satellite) |
| FIRST | Faint Images of the Radio Sky at Twenty-cm (Survey with NRAO's VLA) |
| FITS | Flexible Image Transport System |
| FTP | File Transfer Protocol |
| GASP | Guide-Star Astrometric Support Package (ST ScI) |
| GEISHA | Groningen Exportable Infrared System for High-Resolution Analysis |
| GIPSY | Groningen Image Processing System |
| GRO | Gamma-Ray Observatory (Compton Observatory) |
| GSFC | Goddard Space Flight Center |
| GUI | Graphical User Interface |
| HEADS | High Energy Astrophysics Database Service |
| HEAO | High Energy Astrophysics Observatory |
| HEASARC | High Energy Astrophysics Science Archive Research Center |
| Hipparcos | High-Precision Parallax Collecting Satellite |
| HST | Hubble Space Telescope |
| HTML | Hypertext Markup Language |
| IAU | International Astronomical Union |
| ING | Isaac Newton Group (of telescopes on La Palma, see JKT, INT and WHT) |
| INSPEC | INformation Services in Physics, Electrotechnology, Computers and Control |
| INT | Isaac Newton Telescope (2.5-m mirror on La Palma) |
| IP | Internet protocol |
| IPAC | Infrared Processing and Analysis Center |
| IRAF | Image Reduction and Analysis Facility |
| IRAS | Infrared Astronomical Satellite |
| IRAM | Institut de Radioastronomie Millimétrique |
| ISO | Infrared Space Observatory |
| ISSA | Infrared Sky Survey Atlas (IRAS maps) |
| IUE | International Ultraviolet Explorer |
| JCMT | James Clerk Maxwell Telescope (Hawaii) |
| JKT | Jakobus Kapteyn Telescope (1-m mirror on La Palma) |
| LANL | Los Alamos National Laboratory |
| LDS | Leicester Database System |
| LEDA | Lyon-Meudon Extragalactic Database |
| LHEA | Laboratory for High Energy Astrophysics |
| LPI | Lunar and Planetary Institute |
| MIDAS | Munich Image Data Analysis System |
| MPE | Max-Planck Institut für Extraterrestrische Physik (Garching) |
| MQQ | multiple quick query |
| NASA | National Aeronautics and Space Administration |
| NCSA | National Center for Supercomputing Applications |
| NED | NASA/IPAC Extragalactic Database |
| NFRA | Netherlands Foundations for Research in Astronomy |
| NGDC | National Geophysical Data Center |
| NMC | NASA's Master Catalog |
| NODIS | NSSDC Online Data and Information Service |
| NONA | NSI On-Line Network Aid |
| NSI | NASA Science Internet |
| NSSDC | National Space Science Data Center |
| NOAO | National Optical Astronomy Observatory |
| NRAO | National Radio Astronomy Observatory |
| NTT | New Technology Telescope (ESO, La Silla) |
| NVSS | NRAO VLA Sky Survey |
| OP | Opacity Project |
| PGC | Principal Galaxy Catalogue (see A&AS 80, 299) |
| POSS | Palomar Observatory Sky Survey |
| PPARC | Particle Physics and Astronomy Research Council (UK), previously SERC |
| PPM | Positions and Proper Motions (Star Catalog) |
| RA | Right ascension |
| RAL | Rutherford Appleton Laboratory |
| RAP | Radio Astronomy Preprints (NRAO service) |
| RECON | REmote CONsole (program by NASA STI to access their data) |
| RGO | Royal Greenwich Observatory (Cambridge, UK) |
| ROE | Royal Observatory Edinburgh (Edinburgh, UK) |
| ROSAT | Röntgensatellit |
| SAO | Smithsonian Astrophysical Observatory |
| SCAR | Starlink Catalog Access and Reporting System |
| SCCA | Statistical Consulting Center for Astronomy |
| SDC | Space Data Center |
| SEL | Space Environment Laboratory |
| SERC | Science and Engineering Research Council (UK), now PPARC |
| SGML | Standard Generalized Markup Language |
| SHS | Solar and Heliospheric Division |
| SIMBAD | Set of Identifications, Measurements, and Bibliography for Astronomical Data |
| SISSA | International School for Advanced Studies, Trieste, Italy |
| SMM | Solar Maximum Mission |
| SPAN | Space Physics Analysis Network |
| SPIN | Searchable Physics Information Notices |
| STARCAT | Space Telescope Archive and Catalog |
| STD | Solar Terrestrial Dispatch |
| ST–ECF | Space Telescope – European Coordinating Facility |
| STELAB | Solar-Terrestrial Environment Laboratory |
| STELAR | STudy of Electronic Literature for Astronomical Research |
| STEP | Space Telescope Exhibited Preprints |
| STEP | Solar-Terrestrial Energy Program |
| STI | NASA's Scientific and Technical Information Program |
| STN | Scientific and Technical information Network |
| ST ScI | Space Telescope Science Institute (Baltimore, USA) |
| SUSI | Superb Seeing Imager (ESO) |
| TCP | Transport Connection Protocol |
| UKIRT | United Kingdom Infrared Telescope (on Hawaii) |
| UKST | United Kingdom Schmidt Telescope (Siding Spring, New South Wales) |
| ULDA | IUE Uniform Low Dispersion Archive |
| URL | Uniform Resource Locator |
| VLA | Very Large Array (Socorro, New Mexico) |
| VLAIS | Very Large Array Information System |
| XTE | X-ray Timing Explorer |
| WAIS | Wide-Area Information Servers |
| WHT | William Herschel Telescope (4.2-m mirror on La Palma) |
| WSRT | Westerbork Synthesis Radio Telescope |
| WWW | World-Wide Web |

Table 1 : Some Major Astronomical Data Analysis Packages available via FTP

| System | Org | Archive Address | e-mail contact |
|---|---|---|---|
| IRAF | NOAO | `iraf.noao.edu` (140.252.1.1) | `iraf@noao.edu` |
| STSDAS | ST ScI | `stsci.edu` (130.167.1.2) | `stsdas@stsci.edu` |
| PROS | CfA | `sao-ftp.harvard.edu` (128.103.42.3) | `pros@cfa.harvard.edu` |
| AIPS | NRAO | `baboon.cv.nrao.edu` (192.33.115.103) | `aipsmail@nrao.edu` |
| MIDAS | ESO | `ftphost.hq.eso.org` (134.171.8.4) | `midas@eso.org` |
| VISTA | Lowell | `lowell.edu` (192.103.11.2) | `vista@lowell.edu` |
| FIGARO | AAO | `aaoepp.aao.gov.au` (130.155.203.64) | `ks@aaoepp.aao.gov.au` |
| PGPLOT | Caltech | `deimos.caltech.edu` (131.215.139.14) | `tjp@deimos.caltech.edu` |
| IDL Astr Lib | GSFC | `idlastro.gsfc.nasa.gov` (128.183.57.82) | `landsman@stars.gsfc.nasa.gov` |